\documentclass[11pt, amssymb, nofootinbib, aps, prd,superscriptaddress,preprintnumbers]{revtex4}

\usepackage[utf8]{inputenc}
\usepackage{graphicx}
\usepackage{xcolor}
\usepackage{amsmath}
\usepackage{bm}
\usepackage{amssymb}
\usepackage{amsbsy}
\usepackage{braket}
\usepackage{tikz} 
\usetikzlibrary{shapes,arrows,positioning,automata,backgrounds,calc,er,patterns,snakes,decorations.pathmorphing,decorations.markings}
\usepackage{tikz-feynman}
\usepackage{lipsum}
\usepackage{comment}
\usepackage{hyperref}
\usepackage[normalem]{ulem}

\newcommand{\X}{{\mathbf x}}
\newcommand{\K}{{\mathbf k}}
\newcommand{\PP}{{\mathbf p}}
\newcommand{\Q}{{\mathbf q}}
\newcommand{\V}{{\mathbf v}}
\newcommand{\A}{{\mathbf a}}
\newcommand{\R}{{\mathbf r}}
\newcommand{\RR}{\hat{\mathbf r}}
\newcommand{\ds}{\displaystyle}
\newcommand{\vphi}{\varphi}

\newcommand{\scalar}{thick,decorate,decoration={zigzag,segment length=1.8mm,amplitude=0.4mm}}
\newcommand{\gravphi}{thick,dashed}
\newcommand{\gravA}{thick,photon}
\newcommand{\gravS}{line width=0.8mm,photon}
\newcommand{\gravSS}{line width=0.3mm,photon,white}

\begin{document}

\title{Binary dynamics to second post-Newtonian order in scalar-tensor
and Einstein-scalar-Gauss-Bonnet gravity from effective field theory}

\author{Gabriel Luz Almeida}
\email{galmeida@ustc.edu.cn}
\affiliation{Interdisciplinary Center for Theoretical Study,\\
University of Science and Technology of China, Hefei, Anhui 230026, China}
\affiliation{Peng Huanwu Center for Fundamental Theory, Hefei, Anhui 230026, China}

\begin{abstract}
Using effective field theory methods, we compute in detail the Lagrangian for the conservative dynamics of compact binary systems, for spinless constituents and in the gravitationally bound case, in massless scalar-tensor (ST) and Einstein-scalar-Gauss-Bonnet gravity (EsGB) to the second post-Newtonian (2PN) order. We employ the Kaluza-Klein parametrization of the metric, and demonstrate that, also in this case, a significant reduction in the number of diagrams is achieved. Still, to the 2PN accuracy, an additional 39 diagrams involving scalar interactions must be added to the pure general relativistic result. Diagrams involving the Gauss-Bonnet term at the formal 1PN order are computed and shown to scale as a 3PN correction when observational constraints are taken into account. We also investigate at which order the additional $\mathcal{O}(G^3)$ topologies contribute. Finally, the results derived in this paper provide an essential step toward the correct derivation of the conservative dynamics at the 3PN level in ST and EsGB theories.
\end{abstract}

\preprint{USTC-ICTS/PCFT-24-10}

\maketitle

\section{Introduction}

The recent observations of gravitational waves (GWs) by the LIGO-Virgo-KAGRA collaboration \cite{TheLIGOScientific:2014jea,TheVirgo:2014hva,KAGRA:2020tym} from the coalescence of compact binary systems \cite{LIGOScientific:2016aoc,LIGOScientific:2018mvr,LIGOScientific:2020ibl,LIGOScientific:2021usb,LIGOScientific:2021djp} have provided us with the unique opportunity to explore gravity in its strong-field regime. The future of GW astrophysics is very promising with the advent of ground-based observatories of third generation, particularly the Einstein telescope \cite{Maggiore:2019uih} and
the space-borne interferometer LISA \cite{LISA:2017pwj}, both currently under construction, and expected to be launched in the next decade. With these detectors, coalescing binary systems made of black holes (BHs) and neutron stars (NSs) will reinforce their role as the best cosmic laboratories not only to estimate with better accuracy the astrophysical properties of the individual objects, but also to perform stringent tests on possible deviations from Einstein's theory of general relativity (GR).

Challenges involved in the prediction of gravitational behavior in both short and long scales, most notably the nonrenormalizability of GR as a quantum field theory and the singularity formation as a classical theory, both arising in the strong-field regime, as well as the cosmological constant problem in the large-scale regime, has motivated a quest for an alternative theory of gravity that extends GR. There exists nowadays a plethora of such extensions, all being byproducts of the violation of one or more assumptions of the Lovelock's theorem \cite{Lovelock:1971yv,Lovelock:1972vz}, with new physics in the gravitational sector being almost inescapably attained by the introduction of additional degrees of freedom. The reader is referred to Ref.~\cite{Berti:2015itd} for a comprehensive introduction to the field. 
   
A well-motivated class of alternative theories one might consider is the so-called Horndeski theories \cite{Horndeski:1974wa}. These comprise the most general extensions of GR in which a single scalar field degree of freedom is coupled to gravity and which leads to second-order equations of motion, thus avoiding the emergence of Ostrogradski instabilities \cite{Woodard:2006nt}. One of the most natural and best-studied case of this class is given by the scalar-tensor (ST) theories (See Refs.~\cite{Fujii:2003pa,Damour:1992we} for reviews), which, although very simple, have a rich phenomenology that include their prediction on the characteristic phenomenom of spontaneous scalarization of NSs \cite{Damour:1993hw,Damour:1996ke}. As it happens to the large shift-symmetric class of Horndeski theories \cite{Hui:2012qt}, black hole solutions in ST theories are the same as those of GR \cite{Hawking:1972qk,Sotiriou:2011dz}. 
For this reason, more recently there has been an increase in interest in the so-called Einstein-scalar-Gauss-Bonnet (EsGB) gravity, since in addition to also being a particular case of the Horndeski theories, it evades the no-hair theorem, hence allowing for the existence of BH solutions that differ from GR's \cite{Sotiriou_2014,Sotiriou_2014b}. This is important from the observational point of view since not only BHs are simpler objects than NSs, but also because most of the GW events detected so far correspond to binary BH coalescences, making these objects the best candidates for the search of GW imprints of deviations from GR.

The EsGB gravity extends ST theories (in the Einstein frame) by introducing a nonminimal coupling of the scalar field to gravity through the Gauss-Bonnet invariant, $\mathcal{G}$, defined by $\mathcal{G} \equiv R_{\mu\nu\rho\sigma}R^{\mu\nu\rho\sigma} - 4 R_{\mu\nu}R^{\mu\nu} + R^2$. Note that, as it is long known, the renormalizability of GR is achieved once operators quadratic in the curvature are added to the Einstein-Hilbert action \cite{Stelle:1976gc}, in which, in particular, the Gauss-Bonnet invariant is the only of such combination of quadratic curvature terms that yields second-order equations of motion. Nevertheless, when one considers such a coupling, it is natural to couple it to an scalar field, since, alone, the Gauss-Bonnet invariant is a topological term in four dimensions. In addition to such desirable features, the EsGB gravity is also well-motivated as it naturally arises in the low-energy limit of heterotic string theory \cite{Gross:1986mw}.

In this paper, we are mostly concerned about the second post-Newtonian (2PN) conservative dynamics of compact binaries in EsGB gravity, in the spinless case, to be computed via effective field theory (EFT) methods, where the nonrelativistic general relativity (NRGR) framework of Ref.~\cite{Goldberger:2004jt} is extended to encompass the particular beyond GR theory, which, again, have ST theories as a particular case. Computation of the conservative sector of the binary dynamics is crucial for GW modeling, to either construct simpler orbital phase waveforms or for the modeling of the full inspiral-merger-ringdown waveform by means of, e.g., the effective-one-body approach; See Ref.~\cite{Julie:2022qux} for an application of the latter in the context of ST and EsGB theories. 

In the NRGR framework, the dynamics of compact binary systems is split into two well-separated momentum regions, the so-called near zone, controlled by the potential modes, and the far zone, controlled by the radiation modes, with modes varying on scales $(k_0, |\K|)_{\rm pot}\simeq (v/r, 1/r)$ and $(k_0 , |\K|)_{\rm rad}\simeq (v/r, v/r)$ respectively, $v$ being the relative velocity and $r$ the orbital separation. Such splitting is well defined only in the nonrelativistic limit, $v\ll 1$ (in units where $c=1$), since it enables a hierachy of scales to take place in the gravitational bound problem. Namely, because of the virial theorem, $v^2\sim G M/r$, where $M$ is the total mass of the system, we have that the typical radius of the source $r_s\sim GM$, the orbital separation $r$, and the wavelength of the gravitational radiation $\lambda\sim r/v$, get related by $r_s \ll r \ll \lambda$, hence not only allowing for the splitting between potential and radiation modes, but also making manifest a perturbative expansion in powers of $v$ in each of these momentum regions. It turn, this expansion defines the post-Newtonian approximation, in which the $n$PN order corresponds to the $\mathcal{O}(v^{2n})$ correction to the Newtonian solution.

In the last two decades, the different, yet complementary, methods to perform computation of the pure GR two-body dynamics in the post-Newtonian approximation
(the traditional \cite{Blanchet:2013haa} and EFT \cite{Goldberger:2007hy,Foffa:2013qca,Porto:2016pyg,Levi:2018nxp,Sturani:2021ucg}) have walked hand in hand, and a lot has been learned thus far. In particular, we have witnessed the attainment of the complete 4PN conservative dynamics in the spinless case, both from EFT \cite{Foffa:2019rdf,Foffa:2019yfl} and traditional computations
\cite{Jaranowski:2013lca,Bini:2013zaa,Damour:2014jta,Jaranowski:2015lha,Damour:2016abl,Bernard:2015njp,Bernard:2016wrg}, which involve not only the contribution of potential modes but also the inclusion of the tail effect, which is a process of a radiative nature described in the far zone. Important progress toward a consistent understanding and completion of the 5PN order dynamics have also been accomplished, but most particularly through EFT methods \cite{Almeida:2021xwn,Blumlein:2021txe,Almeida:2022jrv,Almeida:2023yia}. In ST theories, on the other hand, the state-of-the-art computation finds itself in the 3PN order, with 1PN order computed by Damour and Esposito-Farese in \cite{TDamour_1992}, the 2PN by Mirshekari and Will in \cite{Mirshekari:2013vb}, and finally the 3PN order by Bernard in \cite{Bernard:2018hta,Bernard:2018ivi}, and progress in the EsGB gravity standing formally at the 1PN order \cite{BertiJulie2019}. The 1PN order dynamics in ST theories have also been studied within EFT methods in \cite{Kuntz:2019zef} for a massless scalar and in \cite{Diedrichs:2023foj} for the massive case; See also \cite{Bhattacharyya:2023kbh} for a study involving additional degrees of freedom. In these theories, tail effects start to contribute to the conservative dynamics at the 3PN order (1PN order earlier than in GR), hence, up to the 2PN order, the full account of the conservative sector is left to the near zone.

Thus, the main goal of the present paper is twofold. First, to give start to the program of systematic computation of the conservative binary dynamics in ST and EsGB gravity via EFT methods by first reproducing the 2PN order Lagrangian for spinless objects. This is important not only to provide an independent check of the results obtained so far for ST theories, but also because it serves as a necessary building block toward the 3PN order, naturally paving the way for future higher-order computations, similarly to what was initiated in \cite{Gilmore:2008gq} when the 2PN order was computed in the pure GR case from EFT methods. It is worthwhile mentioning here that, as pointed out in Ref.~\cite{Julie:2022qux}, some of the coefficients of the 3PN Lagrangian computed in \cite{Bernard:2018hta} must be revised. Hence, complementary EFT computations become very relevant toward finding the fully consistent 3PN equations of motion. And second, to study from a field-theoretical point of view the relevance of the Gauss-Bonnet coupling in lower PN orders. To this end, we investigate what the diagram topologies of $\mathcal{O}(G^1)$, $\mathcal{O}(G^2)$, and $\mathcal{O}(G^3)$ in ST theories can tell us regarding the PN order in which we expected to observe BHs hair in the context of the EsGB gravity theory. Needless to say, the detailed computation thought Feynman diagrams will justify once again the efficiency of the EFT framework, which gets further reduced once the Kaluza-Klein decomposition of the metric \cite{Kol:2007bc} is employed.

The paper is structured as follows. In Sec.~\ref{sec:EsGBgravity} we introduce the basic formulation of the ST and EsGB theories and discuss how compact objects are described within the post-Newtonian approximation in these theories. In Sec.~\ref{SecEFT} we briefly review the relevant EFT ideas of NRGR and use them to adapt the formalism to include the new scalar degrees of freedom. In this section, we also determine the diagram topologies that are relevant to the 2PN order. In Secs.~\ref{sec1PNconservative} and \ref{sec2PNconservative} we compute all the diagrams entering the 0PN, 1PN, and 2PN orders in ST theories, respectively. In both sections the reduction of diagrams due to the Kaluza-Klein parametrization is discussed. In the end of Sec.~\ref{sec2PNconservative} we present the final and most important result of the paper, namely the 2PN order Lagrangion for binary systems in ST theories. In sec.~\ref{secGBcoupling0} we compute the leading-order contributions that arise from the Gauss-Bonnet coupling, and investigate at which order they contribute. Finally, in Sec.~\ref{secconclusions0} we present the concluding remarks.

\section{The Einstein-Scalar-Gauss-Bonnet Gravity}\label{sec:EsGBgravity}

EsGB theories depart from GR as they contain a massless scalar as additional degree of freedom that couples nonminimally to gravity through the Gauss-Bonnet invariant. 
In the presence of matter, and written in the Einstein frame, the action for such gravity theories takes the form 
\begin{align}\label{eq:GBaction}
S = S_{\rm EH}[g_{\mu\nu}] + S_{\vphi}[\vphi,g_{\mu\nu}] + S_m[\Psi, \mathcal{A}^2(\vphi)g_{\mu\nu}]\,,
\end{align}
where $S_{\rm EH}$ and $S_{\vphi}$ are, respectively, the Einstein-Hilbert action and the coupling of the scalar field with gravity, given by\footnote{Throughout this paper, we use units in which the speed of light is $c=1$ and we adopt the mostly plus signature for the metric, $\eta_{\mu\nu} = {\rm diag}(-1,1,1,1)$.}
\begin{align}
S_{\rm EH} = &\frac{1}{16\pi G} \int d^4x\,\sqrt{-g} R \,, \label{EHaction0} \\
S_{\vphi} = &\frac{1}{16\pi G} \int d^4x\,\sqrt{-g} \left[ - 2g^{\mu\nu}\partial_\mu \vphi \partial_\nu \vphi + \alpha f(\vphi) \mathcal{G} \right]\,,  \label{phiaction0}
\end{align}
and $S_m$ is the matter coupling to both the scalar and tensor fields.
In these expressions, $g = \det g_{\mu\nu}$ is the metric determinant, $R$ is the Ricci scalar, and $\alpha$ is a coupling constant with dimensions of $({\rm length})^2$. 
Knowledge of the functions $\alpha f$ and $\mathcal{A}$ completely determine the EsGB theory.
The quantity denoted by $\mathcal{G}$ is the Gauss-Bonnet invariant, and is given, in addition to the Ricci scalar, in terms of the Riemann tensor $R_{\mu\nu\rho\sigma}$ and Ricci tensor $R_{\mu\nu}$ by
\begin{equation}
\mathcal{G} = R^{\mu\nu\rho\sigma} R_{\mu\nu\rho\sigma} - 4 R^{\mu\nu} R_{\mu\nu} + R^2\,.
\end{equation} 
The action for the matter, $S_m$, is a function of the matter fields $\Psi$, the latter being minimally coupled to the Jordan metric $\tilde{g}_{\mu\nu} = \mathcal{A}^2(\vphi) g_{\mu\nu}$.
Note that the action in Eq.~\eqref{eq:GBaction} reduces to that of ST theories when $\alpha = 0$. This is also the case if the function $f$ is a constant, as the coupling to the Gauss-Bonnet invariant vanishes, since the term $\int d^Dx\sqrt{-g}\mathcal{G}$ is well known to be a boundary term for $D\le 4$. By further assuming that the scalar field $\vphi$ is constant, GR is then recovered. Thus, Eq.~\eqref{eq:GBaction} can be used to study in a unified way the dynamics of hairy BHs in EsGB, when $\alpha\neq 0$, and neutron stars in ST theories, when $\alpha = 0$.

In the PN formalism, the compact bodies in a binary system are treated as pointlike objects. In this description, when scalar field couplings are included, the particles acquire a field-dependent mass $m_A(\vphi)$, following from the ``skeletonization" procedure \cite{1975ApJEardley}, depending on the scalar field at the individual worldlines $x^\mu_A(s_A)$, for bodies $A=1,2$. Such dependence encodes information on the compact objects' internal structure \cite{BertiJulie2019}, with matter coupling given by point-particle action  
\begin{equation}\label{eq:sourceaction}
S_m \rightarrow S^{\rm pp}_m[g_{\mu\nu},\vphi,\{x_A^\mu\}] = - \sum_A \int m_A(\vphi) d\tau_A\,,
\end{equation}
where $d\tau_A = \sqrt{-g_{\mu\nu} dx_A^\mu dx_A^\nu}$. 
From this action, we then define the ``sensitivities" (or ``strong-field parameters") $\alpha_A^0, \beta_A^0, \beta_A'^0, \dots$, which measure the strength of the coupling of each body to the scalar field, by evaluating the following functions at the scalar field background value $\vphi_0$: 
\begin{align}
\alpha_A \equiv \frac{d\ln m_A(\vphi)}{d\vphi} \,, \qquad
\beta_A  \equiv \frac{d\alpha_A(\vphi)}{d\vphi} \,, \qquad
\beta_A' \equiv \frac{d\beta_A(\vphi)}{d\vphi} \,.
\end{align}
Hence, $\alpha_A^0 = \alpha_A(\vphi_0)$, $\beta_A^0 = \beta_A(\vphi_0)$, and so on. Within the PN approximation, the expression for $m_A(\vphi)$ can be parametrized by its expansion about $\vphi_0$,
\begin{align}
m_A(\vphi) &=  m^0_A \bigg[ 1 + \alpha_A^0 (\vphi - \vphi_0) + \frac12 ((\alpha_A^0)^2 + \beta_A^0)(\vphi - \vphi_0)^2  \nonumber\\
&\quad\,\,+\frac{1}{3!} \left[ \alpha^0_A ((\alpha_A^0)^2 + 3\beta_A^0) + \beta'^{0}_A \right] (\vphi - \vphi_0)^3
\bigg]  + \mathcal{O}((\vphi - \vphi_0)^4)\,.
\end{align}

To make contact with the existing literature, it is convenient to introduce the following ST parameters:
\begin{align}
G_{12} &= G (1 + \alpha_1^0 \alpha_2^0)\,, \label{defG12}\\
\bar{\gamma}_{12} &= -\frac{2\alpha_1^0\alpha_2^0}{1+\alpha_1^0\alpha_2^0}\,, \qquad
\bar{\beta}_{1} = \frac{1}{2} \frac{\beta_1^0 (\alpha_2^0)^2}{(1+\alpha_1^0\alpha_2^0)^2}\,, \label{STparameters2}\\
\delta_1 &= \frac{(\alpha_1^0)^2}{(1+\alpha_1^0\alpha_2^0)^2}\,, \qquad
\epsilon_1 = \frac{\beta_1^{'0}(\alpha_2^0)^3}{(1+\alpha_1^0\alpha_2^0)^3}\,, \qquad
\zeta_{12} = \frac{\beta_1^{0}\beta_2^{0}\alpha_1^0\alpha_2^0}{(1+\alpha_1^0\alpha_2^0)^3}\,,\label{STparameters3}
\end{align}
in addition to $\bar{\beta}_2, \delta_2$, and $\epsilon_2$, which are obtained from $\bar{\beta}_1, \delta_1$, and $\epsilon_1$, respectively, by performing the simple exchange $1\leftrightarrow 2$. As we will see, the parameter $G_{12}$ is present already at the Newtonian level, with sensitivities $\alpha^0_1$ and $\alpha^0_2$ playing a role of ``renormalizing" the Newton constant. Meanwhile, the parameters $\bar{\gamma}_{12}$ and $\bar{\beta}_A$ arise at the 1PN order, while parameters $\delta_A$, $\epsilon_A$, and $\zeta_{12}$ start to appear just at the 2PN order.

As mentioned before, the state-of-the-art computation of the conservative dynamics of compact binaries in ST theories lies in the 3PN order, with orders 1PN, 2PN, and 3PN computed by Damour and Esposito-Farèse in \cite{TDamour_1992}, Mirshekari and Will in \cite{Mirshekari:2013vb}, and Bernard in \cite{Bernard:2018hta,Bernard:2018ivi}, respectively. However, in these references, the ST theories were formulated in terms of a Jordan frame, and hence comparison of our Einstein-frame results must be carried out carefully. An explicit translation from one frame to the other can be found in \cite{Julie:2022qux}, where the authors presented the complete 3PN results in the Einstein frame, written in terms of the parameter in Eqs.~\eqref{defG12}, \eqref{STparameters2}, and \eqref{STparameters3}.

Now, before presenting the EFT framework adapted to include the new scalar degree of freedom, let us introduce some important notation that will be used throughout the text: we denote by $\X_A$ the spatial position of body $A$, $\V_A = \dot{\X}_A = d\X_A/dt$ its velocity, and $\A_A = \dot{\V}_A$ its acceleration. The orbital separation between the two compact objects in the bound system is denoted by $r = |\X_1-\X_2|$, and $\RR = (\X_1 - \X_2)/r$.

\section{EFT Framework}\label{SecEFT}

\subsection{Basic setup}

In this section, we adapt the EFT framework developed by W. Goldberger and I. Rothstein, originally devised to implement PN computations within the relativistic (bound) two-body problem in GR,
to accommodate the new scalar degree of freedom in scalar-tensor and sGB theories.
In this EFT description, perturbative calculations are performed by considering fluctuations of the metric around flat Minkowski spacetime, which is usually written in a Lorentz covariant form as 
\begin{equation}
g_{\mu\nu} = \eta_{\mu\nu} + h_{\mu\nu}\,,
\end{equation}
where $h_{\mu\nu}$ represents the metric fluctuations, and now supplemented by the scalar field counterpart
\begin{equation}
\vphi = \vphi_0 + \delta\vphi\,,
\end{equation} 
with $\delta\vphi$, the scalar fluctuations.
In this paper, we work in $d+1$ spacetime dimensions, so that dimensional regularization can be employed, and, for the metric perturbation, we use the Kaluza-Klein decomposition introduced in Ref.~\cite{Kol:2007bc}, in which the metric $g_{\mu\nu}$ is parametrized by a scalar $\phi$, a spatial vector $A_i$ and a symmetric rank-2 spatial tensor $\sigma_{ij}$, that, together, account for the original $(d+1)(d+2)/2$ (ten in $d=3$) degrees of freedom present in $g_{\mu\nu}$. In $d+1$ dimensions, the metric in this decomposition reads  
\begin{eqnarray}\label{kkdecomposition}
g_{\mu\nu} = e^{2\phi/\Lambda}
\left(
\begin{array}{cc}
  -1 & \dfrac{A_i}\Lambda\\
  \dfrac{A_j}\Lambda & \quad e^{-c_d\phi/\Lambda}\left(\delta_{ij}+\dfrac{\sigma_{ij}}\Lambda\right) - \dfrac{A_iA_j}{\Lambda^2}
\end{array}
\right)  \,,
\end{eqnarray}
where we have normalized the fields $\phi, A_i, \sigma_{ij}$ by $\Lambda \equiv (32\pi G)^{-1/2}$, and $c_d \equiv 2(d-1)/(d-2)$. 
The advantage of using such a parametrization is that the propagators for the fields $\phi, A_i, \sigma_{ij}$ get decoupled (i.e., there are no propagator mixing these fields), which ultimately entails a considerable reduction in the number of Feynman diagrams one should account for at each perturbative order. 

This decomposition, which naturally employs a splitting between space and time, is well suited for PN calculations, as the perturbative expansion takes a nonrelativistic form in this formalism.  
Within the PN formalism, the binary dynamics of compact objects (notably BHs and NSs) inspiraling around each other in a bound state can be formulated in terms of an EFT since it presents a clear separation of scales. In particular, because of virial theorem, which relates the relative velocity of the two-body system $v$ to Newton's constant $G$ through $v^2 \sim GM/r$, where $M=m_1+m_2$ is the total mass and $r$ is the radial separation between the two objects, we have $v^2 \sim r_s /r$, assuming the two masses to be of the same order, where $r_s=2G M$ is the Schwarzschild radius of the compact object. Then, because the wavelength of the gravitational radiation emitted by the system scales as $\lambda \sim r/v$, we have $r_s \sim r v^2 \sim \lambda v^3$. Hence, in the nonrelativistic regime, $v\ll 1$, we have well-separated length scales in our problem, and the PN approximation is defined as an expansion of the quantities in powers of $v^2$, the 1PN order being the $v^2$ correction to the Newtonian dynamics, and, more generally, the $n$PN order correcting the Newtonian motion by terms of order $(v^2)^n$ beyond the Newtonian solution.  

Because of the hierarchy of scales above, the method of regions can be applied, which then allow us to study the physics at each scale, one at a time. As in the original ``nonrelativistic general relativity" EFT of Ref.~\cite{Goldberger:2004jt}, our starting description is that of a binary system dynamics at scale $r$, where the internal BH (or NS) degrees of freedom have already being integrated out. In this case the effective coupling of gravity and scalar to the compact object's worldlines is given by the point-particle action \eqref{eq:sourceaction}, which, when written in terms of the decomposition \eqref{kkdecomposition}, takes the form
\begin{widetext}
\begin{equation}\label{SppKK}
S_{m}^{\rm pp} = -\sum_A \int dt\, m(\vphi)e^{\phi/\Lambda}\sqrt{ \left( 1 - \V_A^i \frac{A_i}{\Lambda} \right) - e^{-c_d \phi/\Lambda} \left( \delta_{ij} + \frac{\sigma_{ij}}{\Lambda} \right) \V_A^i \V_A^j}\,,
\end{equation}
\end{widetext}
with fields taking values at spacetime points $(t,\X_A(t))$.
For the gravity and scalar dynamics themselves, we consider the action $S_{\rm EH}$ and $S_{\vphi}$ given in Eqs.~\eqref{EHaction0} and \eqref{phiaction0}, which must be supplemented by a gauge-fixing term in the gravitational sector, chosen here to be the one for the harmonic gauge, so that
\begin{align}
S_{\rm EH+GF} &\equiv S_{\rm EH} + S_{\rm GF}  \nonumber\\
              &= 2\Lambda^2 \int d^{d+1}x\,\sqrt{-g} \left( R - \frac{1}{2} \Gamma^\mu \Gamma^\nu g_{\mu\nu} \right)\,,
\end{align}
where $\Gamma^\mu$ is defined in terms of the Christoffel symbol by $\Gamma^\mu\equiv \Gamma^\mu{}_{\alpha\beta} g^{\alpha\beta}$.

Following from the method of regions, we split the metric and scalar field fluctuations as
\begin{equation}\label{methregionsplitting}
h_{\mu\nu} = H_{\mu\nu} + \bar{h}_{\mu\nu} \qquad \text{and} \qquad \delta\vphi = \Phi + \bar{\vphi}\,,
\end{equation}
where $H_{\mu\nu}$ and $\Phi$ represents the potential modes, with momentum scaling as $k^{\mu} \sim (1/r,v/r)$, and $\bar{h}_{\mu\nu}$ and $\bar\vphi$ standing for the radiation modes, in which case the momentum scales as $k^\mu \sim (v/r, v/r)$.
At this point, it is convenient to rescale the scalar field fluctuations according to $\Phi \rightarrow \Phi/\Lambda$, so as to have same dimensions as the gravitational fields $\phi,A_i, \sigma_{ij}$.
Then, since in ST and sGB theories the conservative contributions stemming from processes of radiative nature enter just at the 3PN order, the potential modes alone already fully account for the conservative sector at the 2PN level. Therefore, in studying the conservative dynamics at the 2PN order, we may just set to zero the radiation modes $\bar{h}_{\mu\nu}$ and $\bar{\vphi}$ in Eq.~\eqref{methregionsplitting}. The momentum region given by $k^\mu \sim (1/r, v/r)$ is commonly referred to as the ``near zone".

\subsection{Feynman diagrams and power counting}

In the EFT framework presented above, the effective action $S_{\rm eff}$ that governs the (purely potential) conservative dynamics of compact binaries in scalar-tensor and sGB theories is obtained from
\begin{equation}\label{pathintegralforSTtheories}
e^{i S_{\rm eff}[x_A^\mu]} = \int  \mathcal{D}\phi \mathcal{D}A_i \mathcal{D}\sigma_{ij} \mathcal{D}\Phi\, \exp\left[ i ( S_{\rm EH+GF}[g_{\mu\nu}] + S_{\vphi}[g_{\mu\nu},\Phi] + S_m[g_{\mu\nu},\Phi,{x_A^\mu}] )  \right] \,,
\end{equation}
with $g_{\mu\nu} = \eta_{\mu\nu} + H_{\mu\nu}$.
This path integral provides an effective description of nonpropagating massive worldlines, bound together through classical scalar and gravitational interactions. Feynman rules for propagators, field self-interactions, and couplings to worldlines can then be derived by expanding the right-hand side of this expression in the fields and in powers of $v^2$, and Feynman diagrams can be used to organize the PN perturbative expansion. In particular, by expanding $S_{\rm EH+GF} + S_{\vphi}$ to quadratic order in the fields, the propagators can be easily read off, for each of the fields: the three fields $\phi, A_i, \sigma_{ij}$, of gravity, defined from $H_{\mu\nu} = g_{\mu\nu} - \eta_{\mu\nu}$, with $g_{\mu\nu}$ given by Eq.~\eqref{kkdecomposition}, and the one for the scalar field $\Phi$. We obtain the following: 
\begin{eqnarray}
\begin{array}{lccl}
\ds D_\phi(k) = \frac{1}{c_d} \mathcal{P}(k) \,, &&&
\ds D_\Phi(k) = \frac{1}{4} \mathcal{P}(k)\,,\\
\ds {[}D_A(k){]}_{ij} = - \delta_{ij} \mathcal{P}(k)  \,, &&&
\ds {[}D_\sigma(k){]}_{ijkl} = \tilde{\delta}_{ijkl} \mathcal{P}(k)\,,
\end{array}
\end{eqnarray}
with $\tilde{\delta}_{ijkl} \equiv \delta_{ik} \delta_{jl} + \delta_{il} \delta_{jk} + (2-c_d)\delta_{ij}\delta_{kl}$ and 
\begin{equation}
\mathcal{P}(k) \equiv -\frac{1}{2}\frac{i}{(\K^2-k_0^2)}\,.
\end{equation}

Because of the relation $v^2 \sim GM/r$, and since the Newtonian solution is of $\mathcal{O}(v^2,G^1 v^0)$, an $n$PN correction corresponds to contributions scaling as $G^{n-k+1}(v^2)^k$, $0\le k \le n+1$. Thus, up to the 2PN order, we must consider diagrams with the following scalings:
\begin{center}
\begin{tabular}{c|cccc}
0PN\, &\, $G^0 v^2$\,, & $G^1 v^0$ \\
\hline
1PN\, &\, $G^0 v^4$\,, & $G^1 v^2$\,, & $G^2 v^0$\\
\hline
2PN\, &\, $G^0 v^6$\,, & $G^1 v^4$\,, & $G^2 v^2$\,, & $G^3 v^0$\,
\end{tabular}
\end{center}
In particular, the kinetic $G^0$ contributions can be directly read off from $S_m^{\rm pp}$ in Eq.~\eqref{SppKK} by setting the fields to zero, so that 
\begin{equation}\label{kinetic}
S_{m,\rm kin}^{\rm pp} = - \sum_A m_A^0 \int dt \sqrt{1-\V_A^2}\,.
\end{equation}
Interactions, on the other hand, always carry powers of $G$, which can be conveniently used to organize the PN expansion as we shall see next. 

As previously mentioned, Feynman diagrams can be derived from the path integral in Eq.~\eqref{pathintegralforSTtheories}, and a definite power counting for the PN expansion can be established.	In this EFT, diagrams are composed of two horizontal lines representing the nonpropagating worldlines of each body, which are connected to each other through propagator lines, which in turn are connected to the horizontal lines through worldline couplings, and graviton and scalar self-interaction vertices. In the computation of the classical quantity $S_{\rm eff}$, only fully connected diagrams with no internal loops contribute\footnote{Diagrams involving lines that start and end in the same worldline can also be ommited as they simply provide corrections to the renormalized masses of the compact binary.}\footnote{Internal loops provide $\mathcal{O}(\hbar)$ (and higher) corrections (and hence quantum) to the Newtonian potential; See Ref.~\cite{Sturani:2021ucg}.}. 
In particular, each worldline coupling contributes with $G^{1/2}$, and each graviton/scalar vertex with $n$ external legs contribute with $G^{n/2-1}$. With this power counting, diagrams can be organized in powers of $G$. 

Once the diagram topologies are organized in orders of $G$, powers of $v$ must be counted. There are three origins of appearances of $v$: (i) from worldline couplings, by expanding $S_m^{\rm pp}$ in Eq.~\eqref{SppKK}, (ii) from time derivatives in self-interaction vertices, since $\partial_0 \sim v/r$ in the near zone, and (iii) from the expansion of the propagators. Asymptotic expansion of propagators is an essential part of the method of regions, which, when applied to our framework, becomes\footnote{For an interesting discussion on the choice of boundary conditions ($i\epsilon$ prescriptions) within NRGR the reader is referred to Ref.~\cite{Foffa:2021pkg}.}
\begin{equation}
\frac{1}{\K^2-k_0^2} = \frac{1}{\K^2} \sum_{n=0}^{\infty} \left( \frac{k_0^2}{\K^2}\right)^n\,.
\end{equation} 
In this expression, each power of $k_0^2/\K^2$ scales like $v^2$, hence making manifest the PN expansion in this formalism. 

\begin{figure}[h]
\centering
\begin{tikzpicture}
\begin{feynman}[scale=0.8, transform shape]
\vertex (a2);
\vertex[small,dot,right=1.5cm of a2] (b2) {};
\vertex[right=1.5cm of b2] (c2);
\vertex[below=1.8cm of a2] (d2);
\vertex[small,dot,right=1.5cm of d2] (e2) {};
\vertex[right=1.5cm of e2] (f2);
\vertex[above=0.6cm of b2] (g2);	
\vertex[right=1.2cm of g2] (h2);
\diagram*{
(a2) --[line width=0.8mm,plain] (c2),
(a2) --[line width=0.3mm,plain,white] (c2),
(d2) --[line width=0.8mm,plain] (f2),
(d2) --[line width=0.3mm,plain,white] (f2),
(b2) -- [thick,plain] (e2),
};
\vertex[dot,right=1.475cm of a2] (b2) {};
\vertex[dot,right=1.475cm of d2] (e2) {};
\end{feynman}
\end{tikzpicture}
\caption{Topology for $\mathcal{O}(G^1)$ contributions to $S_{\rm eff}$. In this diagram, the solid vertical line represents a generic propagator, being either one of the four fields present in our EFT.}
\label{fig:topologyG1}
\end{figure}
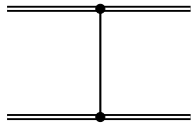

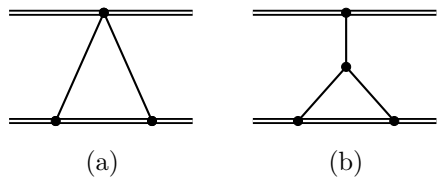
\begin{figure}[h]
\centering
\begin{tikzpicture}
\begin{feynman}[scale=0.8, transform shape]
\vertex (a3);
\vertex[small,dot,right=1.5cm of a3] (b3) {};
\vertex[right=1.5cm of b3] (c3);
\vertex[below=1.8cm of a3] (d3);
\vertex[right=1.5cm of d3] (e3);
\vertex[small,dot,right=0.7cm of d3] (e00) {};
\vertex[small,dot,right=2.3cm of d3] (e000) {};
\vertex[right=1.55cm of e3] (f3);
\vertex[above=0.6cm of b3] (g3);	
\vertex[right=1.2cm of g3] (h3);
\vertex[right=1cm of c3] (a4);
\vertex[small,dot,right=1.5cm of a4] (b4) {};
\vertex[small,dot,below=0.9cm of b4] (g4) {};
\vertex[below=0.45cm of b4] (g40) {};
\vertex[right=1.5cm of g40] (g400);
\vertex[right=1.5cm of b4] (c4);
\vertex[below=1.8cm of a4] (d4);
\vertex[right=1.5cm of d4] (e4);
\vertex[small,dot,right=0.7cm of d4] (e400) {};
\vertex[small,dot,right=2.3cm of d4] (e4000) {};
\vertex[right=1.55cm of e4] (f4);
\vertex[dot,right=1.5cm of a3] (b3) {};
\vertex[dot,right=0.7cm of d3] (e00) {};
\vertex[dot,right=2.3cm of d3] (e000) {};
\diagram*{
(a3) --[line width=0.8mm,plain] (c3),
(a3) --[line width=0.3mm,plain,white] (c3),
(d3) --[line width=0.8mm,plain] (f3),
(d3) --[line width=0.3mm,plain,white] (f3),
(b3) -- [thick,plain] (e00),
(b3) -- [thick,plain] (e000),
(a4) --[line width=0.8mm,plain] (c4),
(a4) --[line width=0.3mm,plain,white] (c4),
(d4) --[line width=0.8mm,plain] (f4),
(d4) --[line width=0.3mm,plain,white] (f4),
(b4) -- [thick,plain] (g4),
(g4) -- [thick,plain] (e400),
(g4) -- [thick,plain] (e4000),
};
\vertex[dot,right=1.5cm of a3] (b3) {};
\vertex[dot,right=0.7cm of d3] (e00) {};
\vertex[dot,right=2.3cm of d3] (e000) {};
\vertex[dot,right=1.475cm of a4] (b4) {};
\vertex[dot,right=0.675cm of d4] (e400) {};
\vertex[dot,right=2.275cm of d4] (e4000) {};
\vertex[dot,below=0.9cm of b4] (g4) {};
\end{feynman}
\begin{feynman}
\vertex (a);
\vertex[below=2cm of a] (b);
\vertex[right=0.89cm of b] (c) {(a)};
\vertex[right=3.24cm of c] (d) {(b)};
\end{feynman}
\end{tikzpicture}
\caption{Topologies for $\mathcal{O}(G^2)$ contributions. }
\label{fig:topologyG2}
\end{figure}

\begin{figure*}
\centering
\begin{tikzpicture}
\begin{feynman}[scale=0.8, transform shape]
\vertex (a);
\vertex[small,dot,right=1.5cm of a] (b) {};
\vertex[small,dot,right=0.7cm of a] (b0);
\vertex[small,dot,right=2.3cm of a] (b00);
\vertex[right=1.5cm of b] (c);
\vertex[below=1.8cm of a] (d);
\vertex[small,dot,right=1.5cm of d] (e) {};
\vertex[small,dot,right=0.7cm of d] (e0) {};
\vertex[small,dot,right=2.3cm of d] (e00) {};
\vertex[right=1.55cm of e] (f);
\vertex[right=1cm of c] (a2);
\vertex[small,dot,white,right=1.5cm of a2] (b2) {};
\vertex[small,dot,right=0.7cm of a2] (b20) {};
\vertex[small,dot,right=2.3cm of a2] (b200) {};
\vertex[right=1.5cm of b2] (c2);
\vertex[below=1.8cm of a2] (d2);
\vertex[small,dot,right=1.5cm of d2] (e2);
\vertex[small,dot,right=0.7cm of d2] (e20) {};
\vertex[small,dot,right=2.3cm of d2] (e200) {};
\vertex[right=1.55cm of e2] (f2);
\vertex[dot,below=0.9cm of b2] (g2);
\vertex[right=1cm of c2] (a3);
\vertex[small,dot,right=1.5cm of a3] (b3) {};
\vertex[right=1.5cm of b3] (c3);
\vertex[below=1.8cm of a3] (d3);
\vertex[small,dot,right=1.5cm of d3] (e3) {};
\vertex[small,dot,right=0.7cm of d3] (e30) {};
\vertex[small,dot,right=2.3cm of d3] (e300) {};
\vertex[right=1.55cm of e3] (f3);
\vertex[dot,below=0.9cm of b3] (g3) {};
\vertex[right=1cm of c3] (a4);
\vertex[small,dot,right=1.5cm of a4] (b4) {};
\vertex[small,dot,right=0.7cm of a4] (b40);
\vertex[small,dot,right=2.3cm of a4] (b400);
\vertex[right=1.5cm of b4] (c4);
\vertex[below=1.8cm of a4] (d4);
\vertex[small,dot,right=1.5cm of d4] (e4) {};
\vertex[small,dot,right=0.7cm of d4] (e40) {};
\vertex[small,dot,right=2.3cm of d4] (e400) {};
\vertex[right=1.55cm of e4] (f4);
\vertex[dot,below=0.9cm of b4] (g4) {};
\vertex[right=1cm of c4] (a5);
\vertex[small,dot,white,right=1.5cm of a5] (b5) {};
\vertex[small,dot,right=0.7cm of a5] (b50) {};
\vertex[small,dot,right=2.3cm of a5] (b500) {};
\vertex[right=1.5cm of b5] (c5);
\vertex[below=1.8cm of a5] (d5);
\vertex[small,dot,right=1.5cm of d5] (e5);
\vertex[small,dot,right=0.7cm of d5] (e50) {};
\vertex[small,dot,right=2.3cm of d5] (e500) {};
\vertex[right=1.55cm of e5] (f5);
\vertex[dot,below=0.9cm of b5] (g5) {};
\vertex[below=1.5cm of d] (a6);
\vertex[small,dot,right=1.5cm of a6] (b6) {};
\vertex[small,dot,right=0.7cm of a6] (b60) {};
\vertex[right=1.5cm of b6] (c6);
\vertex[below=1.8cm of a6] (d6);
\vertex[small,dot,right=1.5cm of d6] (e6);
\vertex[small,dot,right=0.7cm of d6] (e60) {};
\vertex[small,dot,right=2.3cm of d6] (e600) {};
\vertex[right=1.55cm of e6] (f6);
\vertex[dot,below=0.9cm of b6] (g6) {};
\vertex[right=1cm of c6] (a7);
\vertex[small,dot,right=1.5cm of a7] (b7) {};
\vertex[right=1.5cm of b7] (c7);
\vertex[below=1.8cm of a7] (d7);
\vertex[small,dot,right=1.5cm of d7] (e7) {};
\vertex[small,dot,right=0.7cm of d7] (e70) {};
\vertex[small,dot,right=2.3cm of d7] (e700) {};
\vertex[right=1.55cm of e7] (f7);
\vertex[dot,below=0.6cm of b7] (h7) {};
\vertex[dot,above=0.6cm of e7] (h70) {};
\vertex[right=1cm of c7] (a8);
\vertex[small,dot,white,right=1.5cm of a8] (b8) {};
\vertex[small,dot,right=0.7cm of a8] (b80) {};
\vertex[small,dot,right=2.3cm of a8] (b800) {};
\vertex[right=1.5cm of b8] (c8);
\vertex[below=1.8cm of a8] (d8);
\vertex[small,dot,right=1.5cm of d8] (e8);
\vertex[small,dot,right=0.7cm of d8] (e80) {};
\vertex[small,dot,right=2.3cm of d8] (e800) {};
\vertex[right=1.55cm of e8] (f8);
\vertex[dot,below=0.9cm of b80] (g80) {};
\vertex[dot,below=0.9cm of b800] (g800) {};
\vertex[right=1cm of c8] (a9);
\vertex[small,dot,white,right=1.5cm of a9] (b9) {};
\vertex[small,dot,right=0.7cm of a9] (b90) {};
\vertex[small,dot,right=2.3cm of a9] (b900) {};
\vertex[right=1.5cm of b9] (c9);
\vertex[below=1.8cm of a9] (d9);
\vertex[small,dot,white,right=1.5cm of d9] (e9) {};
\vertex[small,dot,right=0.7cm of d9] (e90) {};
\vertex[small,dot,right=2.3cm of d9] (e900) {};
\vertex[right=1.55cm of e9] (f9);
\vertex[dot,below=0.52cm of b9] (h9) {};
\vertex[dot,above=0.52cm of e9] (h90) {};
\diagram*{
(a) --[line width=0.8mm,plain] (c),
(a) --[line width=0.3mm,plain,white] (c),
(d) --[line width=0.8mm,plain] (f),
(d) --[line width=0.3mm,plain,white] (f),
(b) -- [thick,plain] (e0),
(b) -- [thick,plain] (e00),
(b) -- [thick,plain] (e),
(a2) --[line width=0.8mm,plain] (c2),
(a2) --[line width=0.3mm,plain,white] (c2),
(d2) --[line width=0.8mm,plain] (f2),
(d2) --[line width=0.3mm,plain,white] (f2),
(e20) -- [thick,plain] (b20),
(b20) -- [thick,plain] (e200),
(e200) -- [thick,plain] (b200),
(a3) --[line width=0.8mm,plain] (c3),
(a3) --[line width=0.3mm,plain,white] (c3),
(d3) --[line width=0.8mm,plain] (f3),
(d3) --[line width=0.3mm,plain,white] (f3),
(b3) -- [thick,plain] (e3),
(g3) -- [thick,plain] (e30),
(g3) -- [thick,plain] (e300),
(a4) --[line width=0.8mm,plain] (c4),
(a4) --[line width=0.3mm,plain,white] (c4),
(d4) --[line width=0.8mm,plain] (f4),
(d4) --[line width=0.3mm,plain,white] (f4),
(b4) -- [thick,plain] (e4),
(b4) -- [thick,plain] (e400),
(g4) -- [thick,plain] (e40),
(a5) --[line width=0.8mm,plain] (c5),
(a5) --[line width=0.3mm,plain,white] (c5),
(d5) --[line width=0.8mm,plain] (f5),
(d5) --[line width=0.3mm,plain,white] (f5),
(b50) -- [thick,plain] (e500),
(b500) -- [thick,plain] (e50),
(a6) --[line width=0.8mm,plain] (c6),
(a6) --[line width=0.3mm,plain,white] (c6),
(d6) --[line width=0.8mm,plain] (f6),
(d6) --[line width=0.3mm,plain,white] (f6),
(b60) -- [thick,plain] (e60),
(b6) -- [thick,plain] (g6),
(g6) -- [thick,plain] (e60),
(g6) -- [thick,plain] (e600),
(a7) --[line width=0.8mm,plain] (c7),
(a7) --[line width=0.3mm,plain,white] (c7),
(d7) --[line width=0.8mm,plain] (f7),
(d7) --[line width=0.3mm,plain,white] (f7),
(b7) -- [thick,plain] (e7),
(h70) -- [thick,plain] (e70),
(h7) -- [thick,plain] (e700),
(a8) --[line width=0.8mm,plain] (c8),
(a8) --[line width=0.3mm,plain,white] (c8),
(d8) --[line width=0.8mm,plain] (f8),
(d8) --[line width=0.3mm,plain,white] (f8),
(b80) -- [thick,plain] (e80),
(b800) -- [thick,plain] (e800),
(g80) -- [thick,plain] (g800),
(a9) --[line width=0.8mm,plain] (c9),
(a9) --[line width=0.3mm,plain,white] (c9),
(d9) --[line width=0.8mm,plain] (f9),
(d9) --[line width=0.3mm,plain,white] (f9),
(b90) -- [thick,plain] (h9),
(b900) -- [thick,plain] (h9),
(h9) -- [thick,plain] (h90),
(h90) -- [thick,plain] (e90),
(h90) -- [thick,plain] (e900),
};
\vertex[dot,right=1.475cm of a] (bx) {};
\vertex[dot,right=1.475cm of d] (e) {};
\vertex[dot,right=0.675cm of d] (e0) {};
\vertex[dot,right=2.275cm of d] (e00) {};
\vertex[dot,right=0.675cm of a2] (b20) {};
\vertex[dot,right=2.275cm of a2] (b200) {};
\vertex[dot,right=0.675cm of d2] (e20) {};
\vertex[dot,right=2.275cm of d2] (e200) {};
\vertex[dot,right=1.475cm of a3] (b3x) {};
\vertex[dot,right=1.475cm of d3] (e3) {};
\vertex[dot,right=0.675cm of d3] (e30) {};
\vertex[dot,right=2.275cm of d3] (e300) {};
\vertex[dot,right=1.475cm of a4] (b4x) {};
\vertex[dot,right=1.475cm of d4] (e4) {};
\vertex[dot,right=0.675cm of d4] (e40) {};
\vertex[dot,right=2.275cm of d4] (e400) {};
\vertex[dot,right=0.675cm of a5] (b50) {};
\vertex[dot,right=2.275cm of a5] (b500) {};
\vertex[dot,right=0.675cm of d5] (e50) {};
\vertex[dot,right=2.275cm of d5] (e500) {};
\vertex[dot,right=1.475cm of a6] (b6x) {};
\vertex[dot,right=0.675cm of a6] (b60) {};
\vertex[dot,right=0.675cm of d6] (e60) {};
\vertex[dot,right=2.275cm of d6] (e600) {};
\vertex[dot,right=1.475cm of a7] (b7x) {};
\vertex[dot,right=1.475cm of d7] (e7) {};
\vertex[dot,right=0.675cm of d7] (e70) {};
\vertex[dot,right=2.275cm of d7] (e700) {};
\vertex[dot,right=0.675cm of a8] (b80x) {};
\vertex[dot,right=2.275cm of a8] (b800) {};
\vertex[dot,right=0.675cm of d8] (e80) {};
\vertex[dot,right=2.275cm of d8] (e800) {};
\vertex[dot,right=0.675cm of a9] (b90x) {};
\vertex[dot,right=2.275cm of a9] (b900) {};
\vertex[dot,right=0.675cm of d9] (e90) {};
\vertex[dot,right=2.275cm of d9] (e900) {};
\end{feynman}
\begin{feynman}
\vertex (a);
\vertex[below=2cm of a] (b);
\vertex[right=0.89cm of b] (c) {(a)};
\vertex[right=3.24cm of c] (d) {(b)};
\vertex[right=3.24cm of d] (e) {(c)};
\vertex[right=3.24cm of e] (f) {(d)};
\vertex[right=3.24cm of f] (f2) {(e)};
\vertex[below=2.625cm of a] (a2);
\vertex[below=2cm of a2] (b);
\vertex[right=0.89cm of b] (c) {(f)};
\vertex[right=3.24cm of c] (d) {(g)};
\vertex[right=3.24cm of d] (e) {(h)};
\vertex[right=3.24cm of e] (f) {(i)};
\end{feynman}
\end{tikzpicture}
\caption{Topologies for $\mathcal{O}(G^3)$ contributions. }
\label{fig:topologyG3}
\end{figure*}
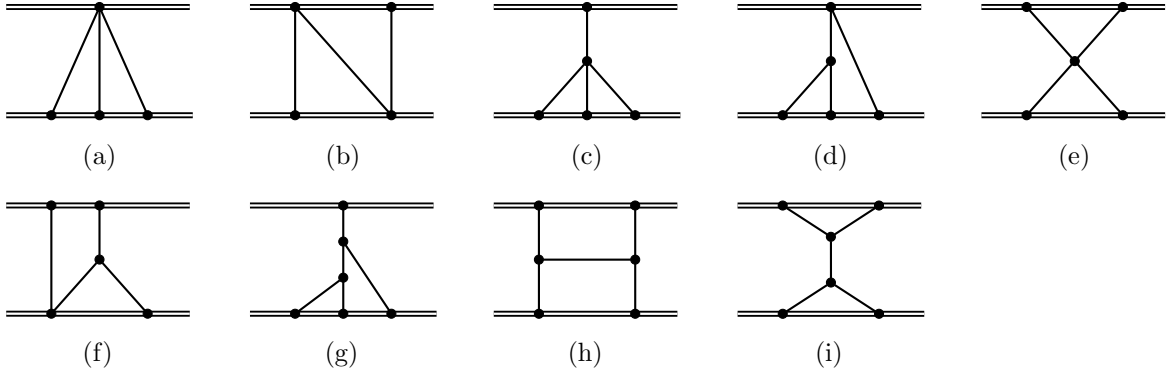

The topologies corresponding to the $G^1$, $G^2$, and $G^3$ contributions entering the 0PN, 1PN and 2PN orders are displayed in Figs.~\ref{fig:topologyG1}, \ref{fig:topologyG2}, and \ref{fig:topologyG3}.
For the worldline coupling of order $v^0$, the diagram topology in Fig.~\ref{fig:topologyG1} reproduces the 0PN (Newtonian) level, while higher powers in $v$, coming either from worldline couplings or from propagator insertions, contribute to higher PN orders. 
Diagrams with the topologies presented in Fig.~\ref{fig:topologyG2} start to contribute to the binary system dynamics at the 1PN order. The three vertex in (b) can be either the pure GR vertex, made of three gravitons, the mixed graviton-scalar-scalar vertex in the ST and sGB theories, or yet the graviton-graviton-scalar interaction present exclusively in sGB theories.

Finally, the diagram topologies of Fig.~\ref{fig:topologyG3} start to contribute to the conservative binary dynamics at the 2PN order. Diagrams with four-vertex interactions are present not only in GR, from its four-graviton vertex, but also in ST and sGB theories, through the couplings of graviton-graviton-scalar-scalar, and graviton-graviton-graviton-scalar interactions, the latter being present uniquely in sGB theories.
As we will show explicitly in Sec.~\ref{sec2PNconservative} in the case of ST theories, the diagram topologies (c), (d), (e), and (f) turn out to be vanishing when the Kaluza-Klein decomposition is employed, justifying the choice of this metric decomposition for computing PN corrections to the conservative binary dynamics from EFT methods. This follows similarly to the pure GR case of Ref.~\cite{Gilmore:2008gq}, where the use of the Kaluza-Klein decomposition is strongly advocated.

\section{Conservative Binary Dynamics to the 1PN Order in ST Theories}\label{sec1PNconservative}

In this section, we start to compute the conservative dynamics of compact binaries, by computing PN corrections to the Lagrangian, restricting ourselves, at this stage, to the ST case, i.e., by setting $\alpha = 0$ in Eq.~\eqref{phiaction0}. We then subsequently complete this result in Sec.~\ref{secGBcoupling0}. by considering the additional contributions stemming from the Gauss-Bonnet coupling, whose results must be added to the ones of this section for the full account of the dynamics in generic sGB gravity theories. 

In this case, for corrections up to the 1PN order, one needs to expand the matter source action $S^{\rm pp}_m$ \eqref{eq:sourceaction} to second order in the fields $\Phi$ and $H_{\mu\nu}$ and to $\mathcal{O}(v^2)$. Nevertheless, to capture just the new contributions due to the coupling of the scalar field to gravity, one just needs to expand the matter action up to $\mathcal{O}(v^0 H^1)$. We thus have
\begin{widetext}
\begin{equation}
S_m^{\rm pp} = -\sum_A m_A^0 \int dt\,  \left[ 1 + \alpha_A^0 \frac{\Phi}{\Lambda} + \frac12 ((\alpha_A^0)^2 + \beta_A^0)\frac{\Phi^2}{\Lambda^2} \right]\times
\left( 1-\frac{\V_A^2}{2} + \frac{\phi}{\Lambda} \right) \,.
\end{equation}
\end{widetext}
In Figs.~\ref{fig:topologyG1} and \ref{fig:topologyG2}, we display the topologies that enter the zeroth and first PN order corrections.

\subsection{The 0PN order}

\begin{figure}[h]
\centering
\begin{tikzpicture}
\begin{feynman}[scale=0.8, transform shape]
\vertex (a2);
\vertex[small,dot,right=1.5cm of a2] (b2) {};
\vertex[right=1.5cm of b2] (c2);
\vertex[below=1.8cm of a2] (d2);
\vertex[small,dot,right=1.5cm of d2] (e2) {};
\vertex[right=1.5cm of e2] (f2);
\vertex[above=0.6cm of b2] (g2);	
\vertex[right=1.2cm of g2] (h2);
\vertex[below=2.5cm of b2] (y) {};
\vertex[right=1cm of c2] (a3);
\vertex[small,dot,right=1.5cm of a3] (b3) {};
\vertex[right=1.5cm of b3] (c3);
\vertex[below=1.8cm of a3] (d3);
\vertex[small,dot,right=1.5cm of d3] (e3) {};
\vertex[right=1.5cm of e3] (f3);
\vertex[above=0.6cm of b3] (g3);	
\vertex[right=1.2cm of g3] (h3);
\vertex[below=2.5cm of b3] (y3) {};
\diagram*{
(a2) --[line width=0.8mm,plain] (c2),
(a2) --[line width=0.3mm,plain,white] (c2),
(d2) --[line width=0.8mm,plain] (f2),
(d2) --[line width=0.3mm,plain,white] (f2),
(b2) -- [\gravphi] (e2),
(a3) --[line width=0.8mm,plain] (c3),
(a3) --[line width=0.3mm,plain,white] (c3),
(d3) --[line width=0.8mm,plain] (f3),
(d3) --[line width=0.3mm,plain,white] (f3),
(b3) -- [\scalar] (e3),
};
\vertex[dot,right=1.475cm of a2] (b2) {};
\vertex[dot,right=1.475cm of d2] (e2) {};
\vertex[above=0.4cm of b2] (v2) {$v^0$};
\vertex[below=0.4cm of e2] (v2) {$v^0$};
\vertex[dot,right=1.475cm of a3] (b3) {};
\vertex[dot,right=1.475cm of d3] (e3) {};
\vertex[above=0.4cm of b3] (v2) {$v^0$};
\vertex[below=0.4cm of e3] (v2) {$v^0$};
\end{feynman}
\begin{feynman}
\vertex (a);
\vertex[below=2.3cm of a2] (b);
\vertex[right=0.89cm of b] (c) {(a)};
\vertex[right=3.24cm of c] (d) {(b)};
\end{feynman}
\end{tikzpicture}
\caption{Diagrams of $\mathcal{O}(G^1)$ contributing to the 0PN order. In diagram (a), the dashed line represents a gravitational $\phi$ propagator and yields the Newtonian potential. Diagram (b) provides a scalar field contribution, with a zigzag line representing the scalar $\Phi$ mode.}
\label{fig:digs0pn}
\end{figure}
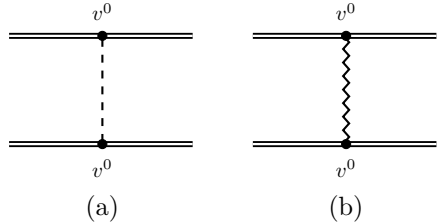

At this level, only the two diagrams of Fig.~\ref{fig:digs0pn} contribute. In particular, diagram (a) presents only one gravitational mode (from a $\phi$ propagator, represented by a dashed line), hence being already present in GR, and was computed for the first time through EFT methods in Ref.~\cite{Goldberger:2004jt}. For completeness, we reproduce this result below. As for diagram (b), which involves one scalar mode (represented by a zigzag line), we compute it here using the Feynman rules determined from the framework presented in Sec.~\ref{SecEFT}, whose results are given by 
\begin{align}
L_{\rm 0PN,(a)} &= \frac{G m_1^0 m_2^0}{r}\,, \label{zeroPNa} \\
L_{\rm 0PN,(b)} &= \frac{G m_1^0 m_2^0 \alpha_1^0 \alpha_2^0}{r} \label{zeroPNb}\,. 
\end{align}
Then, adding these results to the kinetic term given by Eq.~\eqref{kinetic}, and using the notation of Eq.~\eqref{defG12}, we obtain the Lagrangian governing the binary dynamics at the Newtonian level:
\begin{equation}\label{lagrangian0PN}
L_{\rm 0PN} = -m_1^0 -m_2^0 + \frac{1}{2}m_1^0 \V_1^2 + \frac{1}{2}m_2^0 \V_2^2 + \frac{G_{12} m_1^0 m_2^0}{r} \,. 
\end{equation}

In near-zone computations, the following integral, valid more generally in arbitrary $d$ spatial dimensions, is needed: 
\begin{align}\label{intKmaster}
\int_\K \frac{e^{i\K\cdot\X}}{(\K^2)^a} &= \frac{1}{(4\pi)^{d/2}} \frac{\Gamma\left(d/2-a\right)}{\Gamma(a)} \left( \frac{r^2}{4}
\right)^{a-d/2}\,,
\end{align}
where $\int_\K \equiv \int d^d\K/(2\pi)^d$. In particular, in the derivation of the Newtonian contributions in Eqs.~\eqref{zeroPNa} and \eqref{zeroPNb}, the $a = 1$ (with $d=3$) case was used.

\subsection{The 1PN order}

For the 1PN correction to \eqref{lagrangian0PN}, we must consider the three diagram topologies depicted in Figs.~\ref{fig:topologyG1} and \ref{fig:topologyG2}. In particular, for the $\mathcal{O}(G^1)$ topology, we must take into account diagrams that contain $\mathcal{O}(v^1)$ and $\mathcal{O}(v^2)$ terms in either worldline vertices or propagator corrections. The complete set of diagrams contributing  in this case is shown in Fig.~\ref{fig:diags1png1}, where diagrams (a), (b), and (d) are purely GR contributions and have already been computed in \cite{Gilmore:2008gq}, whereas diagrams (c) and (e), which involve a scalar mode, were considered in \cite{Kuntz:2019zef,Bhattacharyya:2023kbh}, and are reproduced here using our notation.

For the $\mathcal{O}(G^2)$ topologies, on the other hand, only diagrams that are absent of $v^2$ corrections in their vertices will contribute to the 1PN order. This is the case since $G^2$ already scales as $\sim v^4$, and hence, any contribution with additional powers of $v^2$ will get pushed to higher PN orders.
In Figs.~\ref{fig:diags1png2} and \ref{fig:diags1png20}, we present all the relevant diagrams for this case. Nevertheless, as we will see below, the diagrams in Fig.~\ref{fig:diags1png20} happen to be all immediately vanishing as a consequence of the Kaluza-Klein decomposition. From the diagrams in Figs.~\ref{fig:diags1png2} and \ref{fig:diags1png20}, only (a) corresponds to pure GR contributions, while the other four are new to the gravitational sector, and are computed below.

\subsubsection{$\mathcal{O}(G^1)$ contributions}

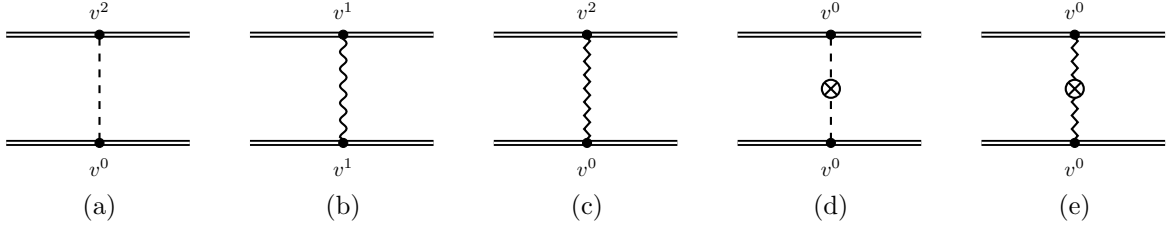
\begin{figure*}
\centering
\begin{tikzpicture}
\begin{feynman}[scale=0.8, transform shape]
\vertex (a2);
\vertex[small,dot,right=1.5cm of a2] (b2) {};
\vertex[right=1.5cm of b2] (c2);
\vertex[below=1.8cm of a2] (d2);
\vertex[small,dot,right=1.5cm of d2] (e2) {};
\vertex[right=1.5cm of e2] (f2);
\vertex[above=0.6cm of b2] (g2);	
\vertex[right=1.2cm of g2] (h2);
\vertex[below=2.5cm of b2] (y) {};
\vertex[right=1cm of c2] (a3);
\vertex[small,dot,right=1.5cm of a3] (b3) {};
\vertex[right=1.5cm of b3] (c3);
\vertex[below=1.8cm of a3] (d3);
\vertex[small,dot,right=1.5cm of d3] (e3) {};
\vertex[right=1.5cm of e3] (f3);
\vertex[above=0.6cm of b3] (g3);	
\vertex[right=1.2cm of g3] (h3);
\vertex[below=2.5cm of b3] (y3) {};
\vertex[right=1cm of c3] (a4);
\vertex[small,dot,right=1.5cm of a4] (b4) {};
\vertex[right=1.5cm of b4] (c4);
\vertex[below=1.8cm of a4] (d4);
\vertex[small,dot,right=1.5cm of d4] (e4) {};
\vertex[right=1.5cm of e4] (f4);
\vertex[above=0.6cm of b4] (g4);	
\vertex[right=1.2cm of g4] (h4);
\vertex[below=2.5cm of b4] (y4) {};
\vertex[right=1cm of c4] (a5);
\vertex[small,dot,right=1.5cm of a5] (b5) {};
\vertex[right=1.5cm of b5] (c5);
\vertex[below=1.8cm of a5] (d5);
\vertex[small,dot,right=1.5cm of d5] (e5) {};
\vertex[right=1.5cm of e5] (f5);
\vertex[above=0.6cm of b5] (g5);	
\vertex[right=1.2cm of g5] (h5);
\vertex[below=2.5cm of b5] (y5) {};
\vertex[right=1cm of c5] (a6);
\vertex[small,dot,right=1.5cm of a6] (b6) {};
\vertex[right=1.5cm of b6] (c6);
\vertex[below=1.8cm of a6] (d6);
\vertex[small,dot,right=1.5cm of d6] (e6) {};
\vertex[right=1.5cm of e6] (f6);
\vertex[above=0.6cm of b6] (g6);	
\vertex[right=1.2cm of g6] (h6);
\vertex[below=2.5cm of b6] (y6) {};
\diagram*{
(a2) --[line width=0.8mm,plain] (c2),
(a2) --[line width=0.3mm,plain,white] (c2),
(d2) --[line width=0.8mm,plain] (f2),
(d2) --[line width=0.3mm,plain,white] (f2),
(b2) -- [\gravphi] (e2),
(a3) --[line width=0.8mm,plain] (c3),
(a3) --[line width=0.3mm,plain,white] (c3),
(d3) --[line width=0.8mm,plain] (f3),
(d3) --[line width=0.3mm,plain,white] (f3),
(b3) -- [\gravA] (e3),
(a4) --[line width=0.8mm,plain] (c4),
(a4) --[line width=0.3mm,plain,white] (c4),
(d4) --[line width=0.8mm,plain] (f4),
(d4) --[line width=0.3mm,plain,white] (f4),
(b4) -- [\scalar] (e4),
(a5) --[line width=0.8mm,plain] (c5),
(a5) --[line width=0.3mm,plain,white] (c5),
(d5) --[line width=0.8mm,plain] (f5),
(d5) --[line width=0.3mm,plain,white] (f5),
(b5) -- [\gravphi] (e5),
(a6) --[line width=0.8mm,plain] (c6),
(a6) --[line width=0.3mm,plain,white] (c6),
(d6) --[line width=0.8mm,plain] (f6),
(d6) --[line width=0.3mm,plain,white] (f6),
(b6) -- [\scalar] (e6),
};
\vertex[dot,right=1.475cm of a2] (b2) {};
\vertex[dot,right=1.475cm of d2] (e2) {};
\vertex[above=0.4cm of b2] (v2) {$v^2$};
\vertex[below=0.4cm of e2] (v2) {$v^0$};
\vertex[dot,right=1.475cm of a3] (b3) {};
\vertex[dot,right=1.475cm of d3] (e3) {};
\vertex[above=0.4cm of b3] (v2) {$v^1$};
\vertex[below=0.4cm of e3] (v2) {$v^1$};
\vertex[dot,right=1.475cm of a4] (b4) {};
\vertex[dot,right=1.475cm of d4] (e4) {};
\vertex[above=0.4cm of b4] (v2) {$v^2$};
\vertex[below=0.4cm of e4] (v2) {$v^0$};
\vertex[dot,right=1.475cm of a5] (b5) {};
\vertex[dot,right=1.475cm of d5] (e5) {};
\vertex[above=0.4cm of b5] (v2) {$v^0$};
\vertex[below=0.4cm of e5] (v2) {$v^0$};
\node[below=0.9cm of b5,dot,scale=1.8,white] (cdot) {};
\node[below=0.9cm of b5,thick,crossed dot] (cdot) {};
\vertex[dot,right=1.475cm of a6] (b6) {};
\vertex[dot,right=1.475cm of d6] (e6) {};
\vertex[above=0.4cm of b6] (v2) {$v^0$};
\vertex[below=0.4cm of e6] (v2) {$v^0$};
\node[below=0.9cm of b6,dot,scale=1.8,white] (cdot) {};
\node[below=0.9cm of b6,thick,crossed dot] (cdot) {};

\end{feynman}
\begin{feynman}
\vertex (a);
\vertex[below=2.3cm of b2] (b) {(a)};
\vertex[below=2.3cm of b3] (b) {(b)};
\vertex[below=2.3cm of b4] (b) {(c)};
\vertex[below=2.3cm of b5] (b) {(d)};
\vertex[below=2.3cm of b6] (b) {(e)};
\end{feynman}
\end{tikzpicture}
\caption{Diagrams with $\mathcal{O}(G^1)$ topology that contribute to the 1PN order. Diagrams (a), (b), and (d) are pure GR contributions, while (c) and (e) are new ones stemming from the scalar field. 
Besides the dashed and zigzag lines already presented, wavy lines represent the propagator for the gravitational $A_i$ mode and crossed circles represent $\mathcal{O}(v^2)$ corrections in the propagator.}
\label{fig:diags1png1}
\end{figure*}

Let us start from the 1PN diagrams with $\mathcal{O}(G^1)$ topology shown in Fig.~\ref{fig:diags1png1}.
Computation of diagrams (c) and (e) is simple, with results given below. Notice that, diagrams (a) and (c) are not symmetric under particle exchange $(1\leftrightarrow 2)$, and therefore the two contributions must be properly taken into account. Again, we also present the results for pure GR, first computed with the use of the Kaluza-Klein fields in Ref.~\cite{Gilmore:2008gq}: 
\begin{align}
L_{\rm 1PN,(a)} &= \frac{3G m_1^0 m_2^0}{2r} (\V_1^2 + \V_2^2)\,, \\
L_{\rm 1PN,(b)} &= -\frac{4G m_1^0 m_2^0}{r} (\V_1\cdot\V_2)\,, \\
L_{\rm 1PN,(c)} &= -\frac{G m_1^0 m_2^0 \alpha_1^0 \alpha_2^0}{2r} (\V_1^2 + \V_2^2)\,, \\
L_{\rm 1PN,(d)} &= \frac{G m_1^0 m_2^0}{2r} \left[ \V_1\cdot\V_2 - (\RR \cdot \V_1 )(\RR \cdot \V_2) \right]\,, \\
L_{\rm 1PN,(e)} &= \frac{G m_1^0 m_2^0 \alpha_1^0 \alpha_2^0}{2r} \left[ \V_1\cdot\V_2 - (\RR \cdot \V_1 )(\RR \cdot \V_2) \right]\,.
\end{align}

\subsubsection{$\mathcal{O}(G^2)$ contributions}

\begin{figure*}
\centering
\begin{tikzpicture}
\begin{feynman}[scale=0.8, transform shape]
\vertex (a);
\vertex[small,dot,right=1.5cm of a] (b) {};
\vertex[small,dot,below=0.9cm of b] (g) {};
\vertex[below=0.45cm of b] (g0) {};
\vertex[right=1.5cm of g0] (g00);
\vertex[right=1.5cm of b] (c);
\vertex[below=1.8cm of a] (d);
\vertex[right=1.5cm of d] (e);
\vertex[small,dot,right=0.7cm of d] (e00) {};
\vertex[small,dot,right=2.3cm of d] (e000) {};
\vertex[right=1.55cm of e] (f);
\vertex[below=2.5cm of b] (y) {};
\vertex[right=1cm of c] (a2);
\vertex[small,dot,right=1.5cm of a2] (b2) {};
\vertex[small,dot,below=0.9cm of b2] (g2) {};
\vertex[below=0.45cm of b2] (g20) {};
\vertex[right=1.5cm of g20] (g200);
\vertex[right=1.5cm of b2] (c2);
\vertex[below=1.8cm of a2] (d2);
\vertex[right=1.5cm of d2] (e2);
\vertex[small,dot,right=0.7cm of d2] (e200) {};
\vertex[small,dot,right=2.3cm of d2] (e2000) {};
\vertex[right=1.55cm of e2] (f2);
\vertex[below=2.5cm of b2] (y2) {};
\vertex[right=1cm of c2] (a3);
\vertex[small,dot,right=1.5cm of a3] (b3) {};
\vertex[small,dot,below=0.9cm of b3] (g3) {};
\vertex[below=0.45cm of b3] (g30) {};
\vertex[right=1.5cm of g30] (g300);
\vertex[right=1.5cm of b3] (c3);
\vertex[below=1.8cm of a3] (d3);
\vertex[right=1.5cm of d3] (e3);
\vertex[small,dot,right=0.7cm of d3] (e300) {};
\vertex[small,dot,right=2.3cm of d3] (e3000) {};
\vertex[right=1.55cm of e3] (f3);
\vertex[below=2.5cm of b3] (y3) {};
\diagram*{
(a) --[line width=0.8mm,plain] (c),
(a) --[line width=0.3mm,plain,white] (c),
(d) --[line width=0.8mm,plain] (f),
(d) --[line width=0.3mm,plain,white] (f),
(b) -- [\gravphi] (g),
(g) -- [\gravphi] (e00),
(g) -- [\gravphi] (e000),
(a2) --[line width=0.8mm,plain] (c2),
(a2) --[line width=0.3mm,plain,white] (c2),
(d2) --[line width=0.8mm,plain] (f2),
(d2) --[line width=0.3mm,plain,white] (f2),
(b2) -- [\gravphi] (g2),
(e200) -- [\scalar] (g2),
(g2) -- [\scalar] (e2000),
(a3) --[line width=0.8mm,plain] (c3),
(a3) --[line width=0.3mm,plain,white] (c3),
(d3) --[line width=0.8mm,plain] (f3),
(d3) --[line width=0.3mm,plain,white] (f3),
(b3) -- [\scalar] (g3),
(g3) -- [\gravphi] (e300),
(g3) -- [\scalar] (e3000),
};
\vertex[dot,right=1.475cm of a] (bx) {};
\vertex[dot,right=0.675cm of d] (e00) {};
\vertex[dot,right=2.275cm of d] (e000) {};
\vertex[dot,below=0.9cm of bx] (g) {};
\vertex[above=0.4cm of bx] (v2) {$v^0$};
\vertex[below=0.4cm of e00] (v2) {$v^0$};
\vertex[below=0.4cm of e000] (v2) {$v^0$};
\vertex[right=0.45cm of g] (z) {};
\vertex[above=0.12cm of z] (v) {$v^0$};
\vertex[dot,right=1.475cm of a2] (b2) {};
\vertex[dot,right=0.675cm of d2] (e200) {};
\vertex[dot,right=2.275cm of d2] (e2000) {};
\vertex[dot,below=0.9cm of b2] (g2) {};
\vertex[above=0.4cm of b2] (v2) {$v^0$};
\vertex[below=0.4cm of e200] (v2) {$v^0$};
\vertex[below=0.4cm of e2000] (v2) {$v^0$};
\vertex[right=0.45cm of g2] (z2) {};
\vertex[above=0.12cm of z2] (v2) {$v^0$};
\vertex[dot,right=1.475cm of a3] (b3) {};
\vertex[dot,right=0.675cm of d3] (e300) {};
\vertex[dot,right=2.275cm of d3] (e3000) {};
\vertex[dot,below=0.9cm of b3] (g3) {};
\vertex[above=0.4cm of b3] (v3) {$v^0$};
\vertex[below=0.4cm of e300] (v3) {$v^0$};
\vertex[below=0.4cm of e3000] (v3) {$v^0$};
\vertex[right=0.45cm of g3] (z3) {};
\vertex[above=0.12cm of z3] (v3) {$v^0$};
\end{feynman}
\begin{feynman}
\vertex[below=2.3cm of b] (bx) {(a)};
\vertex[below=2.3cm of b2] (bx) {(b)};
\vertex[below=2.3cm of b3] (bx) {(c)};
\end{feynman}
\end{tikzpicture}
\caption{Diagrams of $\mathcal{O}(G^2)$ that, in principle, contribute to the 1PN order, but turn out to be vanishing at this level due to the use of the Kaluza-Klein decomposition.}
\label{fig:diags1png20}
\end{figure*}
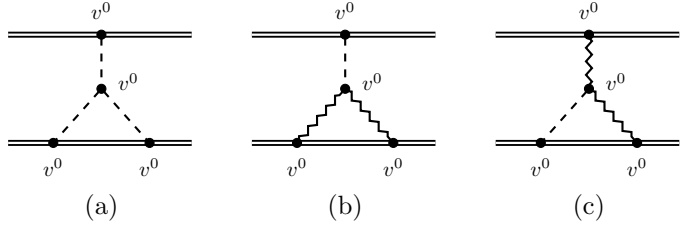

For the $\mathcal{O}(G^2)$, 1PN corrections are derived from the diagrams presented in Figs.~\ref{fig:diags1png2} and \ref{fig:diags1png20}. As previously mentioned, in all these diagrams, only vertices with $\mathcal{O}(v^0)$ are relevant at this order. Then, focusing at the moment on the ``seagull" diagrams of Fig.~\ref{fig:diags1png2}, we see that diagrams (b) and (c) are new, arising from the minimal-coupling of the scalar field to gravity, and are reproduced below (computed also in \cite{Bhattacharyya:2023kbh} using the Kaluza-Klein parametrization and the same gauge choice). From these diagrams, including the pure-GR diagram (a), we obtain the following:
\begin{figure*}
\centering
\begin{tikzpicture}
\begin{feynman}[scale=0.8, transform shape]
\vertex (a3);
\vertex[small,dot,right=1.5cm of a3] (b3) {};
\vertex[right=1.5cm of b3] (c3);
\vertex[below=1.8cm of a3] (d3);
\vertex[right=1.5cm of d3] (e3);
\vertex[small,dot,right=0.7cm of d3] (e300) {};
\vertex[small,dot,right=2.3cm of d3] (e3000) {};
\vertex[right=1.55cm of e3] (f3);
\vertex[above=0.6cm of b3] (g3);	
\vertex[right=1.2cm of g3] (h3);
\vertex[below=2.5cm of b3] (y) {};
\vertex[above=0.4cm of b3] (v) {};
\vertex[right=1cm of c3] (a4);
\vertex[small,dot,right=1.5cm of a4] (b4) {};
\vertex[right=1.5cm of b4] (c4);
\vertex[below=1.8cm of a4] (d4);
\vertex[right=1.5cm of d4] (e4);
\vertex[small,dot,right=0.7cm of d4] (e400) {};
\vertex[small,dot,right=2.3cm of d4] (e4000) {};
\vertex[right=1.55cm of e4] (f4);
\vertex[above=0.6cm of b4] (g4);	
\vertex[right=1.2cm of g4] (h4);
\vertex[below=2.5cm of b4] (y4) {};
\vertex[above=0.4cm of b4] (v4) {};
\vertex[below=0.4cm of e400] (v4) {};
\vertex[below=0.4cm of e4000] (v4) {};
\vertex[right=1cm of c4] (a5);
\vertex[small,dot,right=1.5cm of a5] (b5) {};
\vertex[right=1.5cm of b5] (c5);
\vertex[below=1.8cm of a5] (d5);
\vertex[right=1.5cm of d5] (e5);
\vertex[small,dot,right=0.7cm of d5] (e500) {};
\vertex[small,dot,right=2.3cm of d5] (e5000) {};
\vertex[right=1.55cm of e5] (f5);
\vertex[above=0.6cm of b5] (g5);	
\vertex[right=1.2cm of g5] (h5);
\vertex[below=2.5cm of b5] (y5) {};
\vertex[above=0.4cm of b5] (v5) {};
\vertex[below=0.4cm of e500] (v5) {};
\vertex[below=0.4cm of e5000] (v5) {};
\diagram*{
(a3) --[line width=0.8mm,plain] (c3),
(a3) --[line width=0.3mm,plain,white] (c3),
(d3) --[line width=0.8mm,plain] (f3),
(d3) --[line width=0.3mm,plain,white] (f3),
(b3) -- [\gravphi] (e300),
(b3) -- [\gravphi] (e3000),
(a4) --[line width=0.8mm,plain] (c4),
(a4) --[line width=0.3mm,plain,white] (c4),
(d4) --[line width=0.8mm,plain] (f4),
(d4) --[line width=0.3mm,plain,white] (f4),
(b4) -- [\gravphi] (e400),
(b4) -- [\scalar] (e4000),
(a5) --[line width=0.8mm,plain] (c5),
(a5) --[line width=0.3mm,plain,white] (c5),
(d5) --[line width=0.8mm,plain] (f5),
(d5) --[line width=0.3mm,plain,white] (f5),
(e500) -- [\scalar] (b5),
(b5) -- [\scalar] (e5000),
};
\vertex[dot,right=1.475cm of a3] (b3) {};
\vertex[dot,right=0.675cm of d3] (e300) {};
\vertex[dot,right=2.275cm of d3] (e3000) {};
\vertex[above=0.4cm of b3] (v2) {$v^0$};
\vertex[below=0.4cm of e300] (v2) {$v^0$};
\vertex[below=0.4cm of e3000] (v2) {$v^0$};
\vertex[dot,right=1.475cm of a4] (b4) {};
\vertex[dot,right=0.675cm of d4] (e400) {};
\vertex[dot,right=2.275cm of d4] (e4000) {};
\vertex[above=0.4cm of b4] (v4) {$v^0$};
\vertex[below=0.4cm of e400] (v4) {$v^0$};
\vertex[below=0.4cm of e4000] (v4) {$v^0$};
\vertex[dot,right=1.475cm of a5] (b5) {};
\vertex[dot,right=0.675cm of d5] (e500) {};
\vertex[dot,right=2.275cm of d5] (e5000) {};
\vertex[above=0.4cm of b5] (v5) {$v^0$};
\vertex[below=0.4cm of e500] (v5) {$v^0$};
\vertex[below=0.4cm of e5000] (v5) {$v^0$};
\end{feynman}
\begin{feynman}
\vertex[below=2.3cm of b3] (bx) {(a)};
\vertex[below=2.3cm of b4] (bx) {(b)};
\vertex[below=2.3cm of b5] (bx) {(c)};
\end{feynman}
\end{tikzpicture}
\caption{Diagrams of $\mathcal{O}(G^2)$ that contribute to the 1PN order.}
\label{fig:diags1png2}
\end{figure*}
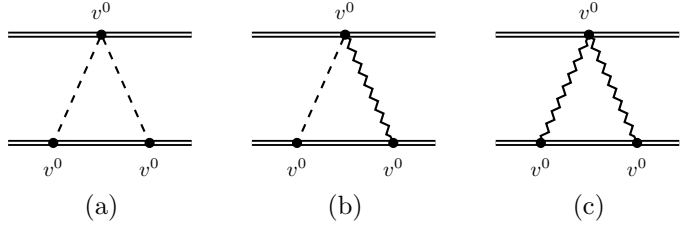
\begin{align}
L_{\rm 1PN,(a)} &= -\frac{G^2 m^0_1 m^0_2 (m^0_1 +  m^0_2)}{2r^2}  \,, \\
L_{\rm 1PN,(b)} &= -\frac{G^2 m^0_1 m^0_2 (m^0_1 +  m^0_2)\alpha_1^0 \alpha_2^0}{r^2}  \,, \\
L_{\rm 1PN,(c)} &= -\frac{G^2 m_1^0 m^0_2}{2r^2} [ m^0_1 ((\alpha_2^0)^2+\beta^0_2) (\alpha^0_1)^2 + m^0_2 ((\alpha_1^0)^2+\beta^0_1) (\alpha^0_2)^2]  \,.
\end{align}
In the derivation of these results, diagrams (a), (b), and (c) of Fig.~\ref{fig:diags1png2} must be supplemented by their symmetric $(1\leftrightarrow 2) $ counterparts.

For the $\mathcal{O}(G^2)$ diagrams in Fig.~\ref{fig:diags1png20}, their results are immediately vanishing at the 1PN order. To see why this happens, consider first the pure GR case of diag.~(a). This diagram is vanishing because the only piece in the three-graviton vertex that contains a $\phi\phi\phi$ interaction, which would then be connected to the leading-order $v^0$ worldline coupling (present only through a $\phi$ coupling in pure GR), contains two time derivatives: $\mathcal{L}_{\phi^3} = -c_d^2 \phi\dot{\phi}^2/\Lambda$ (where $S_{\rm EH+GF} \equiv \int d^{d+1}x\, \mathcal{L}$). Hence, as a consequence, additional powers of $v^2$ are introduced, and this diagram rather contributes to the binary dynamics at the 2PN order. Notice, in particular, that this property is not present when using the metric perturbation $h_{\mu\nu}$ in its Lorentz covariant form, without using the Kaluza-Klein decomposition. 

Similarly, the vanishing of diagrams (b) and (c) occur because the contributions from the metric determinant $\sqrt{-g}$ and inverse $g^{\mu\nu}$ in Eq.~\eqref{phiaction0} to the scalar-scalar-graviton vertex cancel out in the lowest PN order at the level of the action, when the Kaluza-Klein decomposition is employed.
Indeed, at the linear level in $h_{\mu\nu}$, Eq.~\eqref{phiaction0} yields 
\begin{equation}
S_{\vphi} \rightarrow S_{h\Phi\Phi} = 4 \int d^{d+1}x\,  \left( h^{\mu\nu} -\frac12 \eta^{\mu\nu} h  \right) \partial_\mu\Phi \partial_\nu \Phi\,.
\end{equation}
Then, at the 1PN order, only terms with spatial $\mu,\nu$ could contribute, since, again, time derivatives would introduce additional powers of $v$. Thus, using the Kalula-Klein parametrization and neglecting terms containing time derivatives, one gets 
\begin{equation}
S_{h\Phi\Phi} \simeq 4 \int d^{d+1}x\,  \left(  \sigma_{ij} - \frac12 \delta_{ij} \sigma   \right) \partial_i\Phi \partial_j \Phi \,.
\end{equation}
Finally, the lowest-order $\sigma$ coupling contains an additional power of $v^2$ in the worldline action compared to the leading $\mathcal{O}(v^0\phi)$, and hence does not contribute to the 1PN, but only to higher orders.

\subsection{Adding up the contributions}

Finally, we sum all the 1PN contributions computed above, including the 1PN corrections from the kinetic term in Eq.~\eqref{kinetic} and use the notation introduced in Eqs.~\eqref{defG12} and \eqref{STparameters2}. We then arrive in our final result for the 1PN correction to the binary dynamics:
\begin{align}
L_{\rm 1PN} &= \frac{1}{8} m_1^0 \V_1^4 + \frac{1}{8} m_2^0 \V_2^4 + \frac{G_{12}m_1^0 m_2^0}{r} \left[ \frac{3}{2}(\V_1^2+\V_2^2) - \frac{7}{2}(\V_1\cdot\V_2) -  \frac{1}{2}(\RR\cdot\V_1)(\RR\cdot\V_2)+\bar{\gamma}_{12}(\V_1-\V_2)^2
\right] \nonumber\\
& - \frac{G_{12}^2 m_1^0 m_2^0}{2r^2} \left[ m_1^0 (1+2\bar{\beta}_2) + m_2^0 (1+2\bar{\beta}_1)\right]\,.
\end{align}
This result matches Eq.~(4.6) of Ref.~\cite{BertiJulie2019}, computed from traditional methods and in the Einstein frame. Notice that this result differs from the 1PN Lagrangian of Ref.~\cite{Kuntz:2019zef}, computed also from EFT techniques. This happens because a slightly different gauge choice has been used there.

\section{Conservative Binary Dynamics at the 2PN Order in ST Theories}\label{sec2PNconservative}

We move now to the computation of the 2PN order Lagrangian concerning the conservative dynamics of compact binary systems. At this order, we must consider the three sets of diagrams in Figs.~\ref{fig:topologyG1}, \ref{fig:topologyG2}, and \ref{fig:topologyG3}. For the pure GR sector, we take advantage of the results obtained in Ref.~\cite{Gilmore:2008gq}, where the 2PN dynamics was derived through EFT methods, using the same conventions adopted in the present paper (in particular, computed in the harmonic gauge, and using the Kaluza-Klein decomposition). Thus, we restrict ourselves solely to the computation of those diagrams that include scalar field interactions. 
In this case, the relevant worldline couplings we need read as
\begin{widetext}
\begin{align}
S^{\rm pp}_m &= - \sum_{A}\int dt\, m^0_A \left\{ 1 + \alpha_A^0 \frac{\Phi}{\Lambda} + \frac12 ((\alpha_A^0)^2 + \beta_A^0) \frac{\Phi^2}{\Lambda^2} + \frac{1}{3!} \left[ \alpha^0_A ((\alpha_A^0)^2 + 3\beta_A^0) + \beta^{'0}_A \right] \frac{\Phi^3}{\Lambda^3} \right\}
    \nonumber \\
    &\qquad\quad \times \left\{ \left(1-\frac{\V_A^2}{2} -\frac{\V_A^4}{8}\right) + \left[ 1+\frac{(c_d-1)}{2}\V_A^2 \right] \frac{\phi}{\Lambda} 
    - v_A^a \frac{A_a}{\Lambda} -\frac{1}{2} v_A^a v_A^b \frac{\sigma_{ab}}{\Lambda} + \frac{\phi^2}{2\Lambda^2}
    \right\}  \,.
\end{align}
\end{widetext}

\subsection{$\mathcal{O}(G^1)$ contributions}

The relevant diagrams that enter this sector are displayed in Fig.~\ref{diagsG12PNorder}. Then, employing the Feynman rules and performing integration with the help of Eq.~\eqref{intKmaster}, we obtain
\begin{figure*}
\centering
\begin{tikzpicture}
\begin{feynman}[scale=0.8, transform shape]
\vertex (a2);
\vertex[small,dot,right=1.5cm of a2] (b2) {};
\vertex[right=1.5cm of b2] (c2);
\vertex[below=1.8cm of a2] (d2);
\vertex[small,dot,right=1.5cm of d2] (e2) {};
\vertex[right=1.5cm of e2] (f2);
\vertex[above=0.6cm of b2] (g2);	
\vertex[right=1.2cm of g2] (h2);
\vertex[below=2.5cm of b2] (y) {};
\vertex[right=1cm of c2] (a3);
\vertex[small,dot,right=1.5cm of a3] (b3) {};
\vertex[right=1.5cm of b3] (c3);
\vertex[below=1.8cm of a3] (d3);
\vertex[small,dot,right=1.5cm of d3] (e3) {};
\vertex[right=1.5cm of e3] (f3);
\vertex[above=0.6cm of b3] (g3);	
\vertex[right=1.2cm of g3] (h3);
\vertex[below=2.5cm of b3] (y3) {};
\vertex[right=1cm of c3] (a4);
\vertex[small,dot,right=1.5cm of a4] (b4) {};
\vertex[right=1.5cm of b4] (c4);
\vertex[below=1.8cm of a4] (d4);
\vertex[small,dot,right=1.5cm of d4] (e4) {};
\vertex[right=1.5cm of e4] (f4);
\vertex[above=0.6cm of b4] (g4);	
\vertex[right=1.2cm of g4] (h4);
\vertex[below=2.5cm of b4] (y4) {};
\vertex[right=1cm of c4] (a5);
\vertex[small,dot,right=1.5cm of a5] (b5) {};
\vertex[right=1.5cm of b5] (c5);
\vertex[below=1.8cm of a5] (d5);
\vertex[small,dot,right=1.5cm of d5] (e5) {};
\vertex[right=1.5cm of e5] (f5);
\vertex[above=0.6cm of b5] (g5);	
\vertex[right=1.2cm of g5] (h5);
\vertex[below=2.5cm of b5] (y5) {};
\diagram*{
(a2) --[line width=0.8mm,plain] (c2),
(a2) --[line width=0.3mm,plain,white] (c2),
(d2) --[line width=0.8mm,plain] (f2),
(d2) --[line width=0.3mm,plain,white] (f2),
(b2) -- [\scalar] (e2),
(a3) --[line width=0.8mm,plain] (c3),
(a3) --[line width=0.3mm,plain,white] (c3),
(d3) --[line width=0.8mm,plain] (f3),
(d3) --[line width=0.3mm,plain,white] (f3),
(b3) -- [\scalar] (e3),
(a4) --[line width=0.8mm,plain] (c4),
(a4) --[line width=0.3mm,plain,white] (c4),
(d4) --[line width=0.8mm,plain] (f4),
(d4) --[line width=0.3mm,plain,white] (f4),
(b4) -- [\scalar] (e4),
(a5) --[line width=0.8mm,plain] (c5),
(a5) --[line width=0.3mm,plain,white] (c5),
(d5) --[line width=0.8mm,plain] (f5),
(d5) --[line width=0.3mm,plain,white] (f5),
(b5) -- [\scalar] (e5),
};
\vertex[dot,right=1.475cm of a2] (b2) {};
\vertex[dot,right=1.475cm of d2] (e2) {};
\vertex[above=0.4cm of b2] (v2) {$v^2$};
\vertex[below=0.4cm of e2] (v2) {$v^2$};
\vertex[dot,right=1.475cm of a3] (b3) {};
\vertex[dot,right=1.475cm of d3] (e3) {};
\vertex[above=0.4cm of b3] (v2) {$v^4$};
\vertex[below=0.4cm of e3] (v2) {$v^0$};
\vertex[dot,right=1.475cm of a4] (b4) {};
\vertex[dot,right=1.475cm of d4] (e4) {};
\vertex[above=0.4cm of b4] (v2) {$v^2$};
\node[below=0.9cm of b4,dot,scale=1.8,white] (cdot) {};
\node[below=0.9cm of b4,thick,crossed dot] (cdot) {};
\vertex[below=0.4cm of e4] (v2) {$v^0$};
\vertex[dot,right=1.475cm of a5] (b5) {};
\vertex[dot,right=1.475cm of d5] (e5) {};
\vertex[above=0.4cm of b5] (v2) {$v^0$};
\vertex[below=0.4cm of e5] (v2) {$v^0$};
\node[below=0.65cm of b5,dot,scale=1.8,white] (cdot) {};
\node[below=0.65cm of b5,thick,crossed dot] (cdot) {};
\node[below=1.25cm of b5,dot,scale=1.8,white] (cdot) {};
\node[below=1.25cm of b5,thick,crossed dot] (cdot) {};
\end{feynman}
\begin{feynman}
\vertex (a);
\vertex[below=2.3cm of b2] (b) {(i)};
\vertex[below=2.3cm of b3] (b) {(ii)};
\vertex[below=2.3cm of b4] (b) {(iii)};
\vertex[below=2.3cm of b5] (b) {(iv)};
\end{feynman}
\end{tikzpicture}
\caption{Diagrams entering the $\mathcal{O}(G^1)$ sector of the 2PN order dynamics.}
\label{diagsG12PNorder}
\end{figure*}
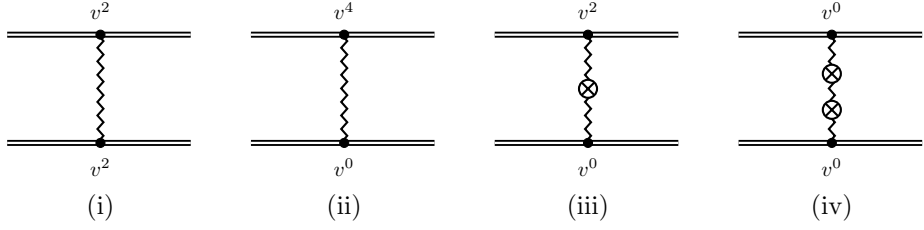
\begin{widetext}
\begin{align}
L_{\rm (i)} & = \frac{G m^0_1 m^0_2 \alpha_1^0\alpha_2^0}{4r} \V_1^2 \V_2^2 \,,\\
L_{\rm (ii)} & = - \frac{G m^0_1 m^0_2 \alpha_1^0\alpha_2^0}{8r} (\V_1^4 + \V_2^4 ) \,,\\
L_{\rm (iii)} & = - \frac{G m^0_1 m^0_2 \alpha_1^0\alpha_2^0}{4r} \Big\{ 2r [ (\RR \cdot \V_2) (\V_1 \cdot \A_1) - (\RR \cdot \V_1) (\V_2\cdot\A_2) ] +(\V_1^2+\V_2^2) [ \V_1\cdot\V_2 - (\RR \cdot \V_1)(\RR \cdot \V_2) ]\Big\} 
 \,,\\
L_{\rm (iv)} &= - \frac{G m^0_1 m^0_2 \alpha_1^0\alpha_2^0}{8r} \Big\{ r^2 [ \A_1\cdot\A_2 + (\RR\cdot\A_1) (\RR\cdot\A_2)] + 2r [ (\RR\cdot\V_1) (\V_1\cdot\A_2) - (\RR\cdot\V_2) (\V_2\cdot\A_1) ] \nonumber\\
&- r [ (\RR\cdot\V_1)^2 (\RR\cdot\A_2) - (\RR\cdot\V_2)^2 (\RR\cdot\A_1) ]  + r [ \V_1^2 (\RR\cdot\A_2) - \V_2^2 (\RR\cdot\A_1) ] +4 (\RR\cdot \V_1)(\RR\cdot \V_2) (\V_1\cdot\V_2) \nonumber\\
&-2(\V_1\cdot\V_2)^2 -3 (\RR\cdot \V_1)^2 (\RR\cdot \V_2)^2 - \V_1^2 \V_2^2  + \V_1^2 (\RR\cdot\V_2)^2 + \V_2^2 (\RR\cdot\V_1)^2 \Big\}\,.
\end{align} 
\end{widetext}

Notice that the contributions $L_{\rm(iii)}$ and $L_{\rm(iv)}$ contain acceleration-dependent terms. This happens because, when written in time domain, every propagator correction (represented by a crossed circle) gives rise to two time derivatives, which then acts on either $\X_1$ or $\X_2$. In particular, since diagram (iv) contains two of such corrections, we observe the presence of a term $\sim \A_1\cdot\A_2$.


\subsection{$\mathcal{O}(G^2)$ contributions}

For the $\mathcal{O}(G^2)$ contributions to the 2PN order, we have the two topologies shown in Fig.~\ref{fig:topologyG2}, (a) and (b). In both cases, the asymmetry under particle exchange allow us to obtain the mirrored contributions by simply making $(1 \leftrightarrow 2)$, taking notice that $\RR \rightarrow -\RR$ under this transformation.

The topology (a) contains nine contributing diagrams, shown in Fig.~\ref{DiagsaG2order2PN}. Computation yields
\begin{figure*}
\centering
\begin{tikzpicture}
\begin{feynman}[scale=0.8, transform shape]
\vertex (a3);
\vertex[small,dot,right=1.5cm of a3] (b3) {};
\vertex[right=1.5cm of b3] (c3);
\vertex[below=1.8cm of a3] (d3);
\vertex[right=1.5cm of d3] (e3);
\vertex[small,dot,right=0.7cm of d3] (e300) {};
\vertex[small,dot,right=2.3cm of d3] (e3000) {};
\vertex[right=1.55cm of e3] (f3);
\vertex[above=0.6cm of b3] (g3);	
\vertex[right=1.2cm of g3] (h3);
\vertex[below=2.5cm of b3] (y) {};
\vertex[above=0.4cm of b3] (v) {};
\vertex[right=1cm of c3] (a4);
\vertex[small,dot,right=1.5cm of a4] (b4) {};
\vertex[right=1.5cm of b4] (c4);
\vertex[below=1.8cm of a4] (d4);
\vertex[right=1.5cm of d4] (e4);
\vertex[small,dot,right=0.7cm of d4] (e400) {};
\vertex[small,dot,right=2.3cm of d4] (e4000) {};
\vertex[right=1.55cm of e4] (f4);
\vertex[above=0.6cm of b4] (g4);	
\vertex[right=1.2cm of g4] (h4);
\vertex[below=2.5cm of b4] (y4) {};
\vertex[above=0.4cm of b4] (v4) {};
\vertex[below=0.4cm of e400] (v4) {};
\vertex[below=0.4cm of e4000] (v4) {};
\vertex[right=1cm of c4] (a5);
\vertex[small,dot,right=1.5cm of a5] (b5) {};
\vertex[right=1.5cm of b5] (c5);
\vertex[below=1.8cm of a5] (d5);
\vertex[right=1.5cm of d5] (e5);
\vertex[small,dot,right=0.7cm of d5] (e500) {};
\vertex[small,dot,right=2.3cm of d5] (e5000) {};
\vertex[right=1.55cm of e5] (f5);
\vertex[above=0.6cm of b5] (g5);	
\vertex[right=1.2cm of g5] (h5);
\vertex[below=2.5cm of b5] (y5) {};
\vertex[above=0.4cm of b5] (v5) {};
\vertex[below=0.4cm of e500] (v5) {};
\vertex[below=0.4cm of e5000] (v5) {};
\vertex[right=1cm of c5] (a6);
\vertex[small,dot,right=1.5cm of a6] (b6) {};
\vertex[right=1.5cm of b6] (c6);
\vertex[below=1.8cm of a6] (d6);
\vertex[right=1.5cm of d6] (e6);
\vertex[small,dot,right=0.7cm of d6] (e600) {};
\vertex[small,dot,right=2.3cm of d6] (e6000) {};
\vertex[right=1.55cm of e6] (f6);
\vertex[above=0.6cm of b6] (g6);	
\vertex[right=1.2cm of g6] (h6);
\vertex[below=2.5cm of b6] (y6) {};
\vertex[above=0.4cm of b6] (v6) {};
\vertex[below=0.4cm of e600] (v6) {};
\vertex[below=0.4cm of e6000] (v6) {};
\vertex[right=1cm of c6] (a7);
\vertex[small,dot,right=1.5cm of a7] (b7) {};
\vertex[right=1.5cm of b7] (c7);
\vertex[below=1.8cm of a7] (d7);
\vertex[right=1.5cm of d7] (e7);
\vertex[small,dot,right=0.7cm of d7] (e700) {};
\vertex[small,dot,right=2.3cm of d7] (e7000) {};
\vertex[right=1.55cm of e7] (f7);
\vertex[above=0.6cm of b7] (g7);	
\vertex[right=1.2cm of g7] (h7);
\vertex[below=2.5cm of b7] (y7) {};
\vertex[above=0.4cm of b7] (v7) {};
\vertex[below=0.4cm of e700] (v7) {};
\vertex[below=0.4cm of e7000] (v7) {};
\vertex[below=1.8cm of d3] (a8);
\vertex[small,dot,right=1.5cm of a8] (b8) {};
\vertex[right=1.5cm of b8] (c8);
\vertex[below=1.8cm of a8] (d8);
\vertex[right=1.5cm of d8] (e8);
\vertex[small,dot,right=0.7cm of d8] (e800) {};
\vertex[small,dot,right=2.3cm of d8] (e8000) {};
\vertex[right=1.55cm of e8] (f8);
\vertex[above=0.6cm of b8] (g8);	
\vertex[right=1.2cm of g8] (h8);
\vertex[below=2.5cm of b8] (y8) {};
\vertex[above=0.4cm of b8] (v8) {};
\vertex[below=0.4cm of e800] (v8) {};
\vertex[below=0.4cm of e8000] (v8) {};
\vertex[right=1cm of c8] (a9);
\vertex[small,dot,right=1.5cm of a9] (b9) {};
\vertex[right=1.5cm of b9] (c9);
\vertex[below=1.8cm of a9] (d9);
\vertex[right=1.5cm of d9] (e9);
\vertex[small,dot,right=0.7cm of d9] (e900) {};
\vertex[small,dot,right=2.3cm of d9] (e9000) {};
\vertex[right=1.55cm of e9] (f9);
\vertex[above=0.6cm of b9] (g9);	
\vertex[right=1.2cm of g9] (h9);
\vertex[below=2.5cm of b9] (y9) {};
\vertex[above=0.4cm of b9] (v9) {};
\vertex[below=0.4cm of e900] (v9) {};
\vertex[below=0.4cm of e9000] (v9) {};
\vertex[right=1cm of c9] (a9x);
\vertex[small,dot,right=1.5cm of a9x] (b9x) {};
\vertex[right=1.5cm of b9x] (c9x);
\vertex[below=1.8cm of a9x] (d9x);
\vertex[right=1.5cm of d9x] (e9x);
\vertex[small,dot,right=0.7cm of d9x] (e9x00) {};
\vertex[small,dot,right=2.3cm of d9x] (e9x000) {};
\vertex[right=1.55cm of e9x] (f9x);
\vertex[above=0.6cm of b9x] (g9x);	
\vertex[right=1.2cm of g9x] (h9x);
\vertex[below=2.5cm of b9x] (y9x) {};
\vertex[above=0.4cm of b9x] (v9x) {};
\vertex[below=0.4cm of e9x00] (v9x) {};
\vertex[below=0.4cm of e9x000] (v9x) {};
\vertex[right=1cm of c9x] (a9y);
\vertex[small,dot,right=1.5cm of a9y] (b9y) {};
\vertex[right=1.5cm of b9y] (c9y);
\vertex[below=1.8cm of a9y] (d9y);
\vertex[right=1.5cm of d9y] (e9y);
\vertex[small,dot,right=0.7cm of d9y] (e9y00) {};
\vertex[small,dot,right=2.3cm of d9y] (e9y000) {};
\vertex[right=1.55cm of e9y] (f9y);
\vertex[above=0.6cm of b9y] (g9y);	
\vertex[right=1.2cm of g9y] (h9y);
\vertex[below=2.5cm of b9y] (y9y) {};
\vertex[above=0.4cm of b9y] (v9y) {};
\vertex[below=0.4cm of e9y00] (v9y) {};
\vertex[below=0.4cm of e9y000] (v9y) {};
\diagram*{
(a3) --[line width=0.8mm,plain] (c3),
(a3) --[line width=0.3mm,plain,white] (c3),
(d3) --[line width=0.8mm,plain] (f3),
(d3) --[line width=0.3mm,plain,white] (f3),
(b3) -- [\scalar] (e300),
(b3) -- [\scalar] (e3000),
(a4) --[line width=0.8mm,plain] (c4),
(a4) --[line width=0.3mm,plain,white] (c4),
(d4) --[line width=0.8mm,plain] (f4),
(d4) --[line width=0.3mm,plain,white] (f4),
(b4) -- [\scalar] (e400),
(b4) -- [\scalar] (e4000),
(a5) --[line width=0.8mm,plain] (c5),
(a5) --[line width=0.3mm,plain,white] (c5),
(d5) --[line width=0.8mm,plain] (f5),
(d5) --[line width=0.3mm,plain,white] (f5),
(e500) -- [\scalar] (b5),
(b5) -- [\gravphi] (e5000),
(a6) --[line width=0.8mm,plain] (c6),
(a6) --[line width=0.3mm,plain,white] (c6),
(d6) --[line width=0.8mm,plain] (f6),
(d6) --[line width=0.3mm,plain,white] (f6),
(e600) -- [\scalar] (b6),
(b6) -- [\gravphi] (e6000),
(a7) --[line width=0.8mm,plain] (c7),
(a7) --[line width=0.3mm,plain,white] (c7),
(d7) --[line width=0.8mm,plain] (f7),
(d7) --[line width=0.3mm,plain,white] (f7),
(e700) -- [\scalar] (b7),
(b7) -- [\gravphi] (e7000),
(a8) --[line width=0.8mm,plain] (c8),
(a8) --[line width=0.3mm,plain,white] (c8),
(d8) --[line width=0.8mm,plain] (f8),
(d8) --[line width=0.3mm,plain,white] (f8),
(e800) -- [\scalar] (b8),
(b8) -- [\gravA] (e8000),
(a9) --[line width=0.8mm,plain] (c9),
(a9) --[line width=0.3mm,plain,white] (c9),
(d9) --[line width=0.8mm,plain] (f9),
(d9) --[line width=0.3mm,plain,white] (f9),
(e900) -- [\scalar] (b9),
(b9) -- [\scalar] (e9000),
(a9x) --[line width=0.8mm,plain] (c9x),
(a9x) --[line width=0.3mm,plain,white] (c9x),
(d9x) --[line width=0.8mm,plain] (f9x),
(d9x) --[line width=0.3mm,plain,white] (f9x),
(e9x00) -- [\scalar] (b9x),
(b9x) -- [\gravphi] (e9x000),
(a9y) --[line width=0.8mm,plain] (c9y),
(a9y) --[line width=0.3mm,plain,white] (c9y),
(d9y) --[line width=0.8mm,plain] (f9y),
(d9y) --[line width=0.3mm,plain,white] (f9y),
(e9y00) -- [\scalar] (b9y),
(b9y) -- [\gravphi] (e9y000),
};
\vertex[dot,right=1.475cm of a3] (b3) {};
\vertex[dot,right=0.675cm of d3] (e300) {};
\vertex[dot,right=2.275cm of d3] (e3000) {};
\vertex[above=0.4cm of b3] (v2) {$v^2$};
\vertex[below=0.4cm of e300] (v2) {};
\vertex[below=0.4cm of e3000] (v2) {};
\vertex[dot,right=1.475cm of a4] (b4) {};
\vertex[dot,right=0.675cm of d4] (e400) {};
\vertex[dot,right=2.275cm of d4] (e4000) {};
\vertex[above=0.4cm of b4] (v4) {};
\vertex[below=0.4cm of e400] (v4) {$v^2$};
\vertex[below=0.4cm of e4000] (v4) {};
\vertex[dot,right=1.475cm of a5] (b5) {};
\vertex[dot,right=0.675cm of d5] (e500) {};
\vertex[dot,right=2.275cm of d5] (e5000) {};
\vertex[above=0.4cm of b5] (v5) {$v^2$};
\vertex[below=0.4cm of e500] (v5) {};
\vertex[below=0.4cm of e5000] (v5) {};
\vertex[dot,right=1.475cm of a6] (b6) {};
\vertex[dot,right=0.675cm of d6] (e600) {};
\vertex[dot,right=2.275cm of d6] (e6000) {};
\vertex[above=0.4cm of b6] (v6) {};
\vertex[below=0.4cm of e600] (v6) {$v^2$};
\vertex[below=0.4cm of e6000] (v6) {};
\vertex[dot,right=1.475cm of a7] (b7) {};
\vertex[dot,right=0.675cm of d7] (e700) {};
\vertex[dot,right=2.275cm of d7] (e7000) {};
\vertex[above=0.4cm of b7] (v7) {};
\vertex[below=0.4cm of e700] (v7) {};
\vertex[below=0.4cm of e7000] (v7) {$v^2$};
\vertex[dot,right=1.475cm of a8] (b8) {};
\vertex[dot,right=0.675cm of d8] (e800) {};
\vertex[dot,right=2.275cm of d8] (e8000) {};
\vertex[above=0.4cm of b8] (v8) {$v^1$};
\vertex[below=0.4cm of e800] (v8) {};
\vertex[below=0.4cm of e8000] (v8) {$v^1$};
\vertex[dot,right=1.475cm of a9] (b9) {};
\vertex[dot,right=0.675cm of d9] (e900) {};
\vertex[dot,right=2.275cm of d9] (e9000) {};
\vertex[above=0.4cm of b9] (v9) {};
\vertex[below=0.4cm of e900] (v9) {};
\vertex[below=0.4cm of e9000] (v9) {};
\vertex[below=0.95cm of b9] (b9p) {};
\node[right=0.42cm of b9p,dot,scale=1.8,white] (cdot) {};
\node[right=0.42cm of b9p,thick,crossed dot] (cdot) {};
\vertex[dot,right=1.475cm of a9x] (b9x) {};
\vertex[dot,right=0.675cm of d9x] (e9x00) {};
\vertex[dot,right=2.275cm of d9x] (e9x000) {};
\vertex[above=0.4cm of b9x] (v9x) {};
\vertex[below=0.4cm of e9x00] (v9x) {};
\vertex[below=0.4cm of e9x000] (v9x) {};
\vertex[below=0.95cm of b9x] (b9xp) {};
\node[right=0.42cm of b9xp,dot,scale=1.8,white] (cdot) {};
\node[right=0.42cm of b9xp,thick,crossed dot] (cdot) {};
\vertex[dot,right=1.475cm of a9y] (b9y) {};
\vertex[dot,right=0.675cm of d9y] (e9y00) {};
\vertex[dot,right=2.275cm of d9y] (e9y000) {};
\vertex[above=0.4cm of b9y] (v9y) {};
\vertex[below=0.4cm of e9y00] (v9y) {};
\vertex[below=0.4cm of e9y000] (v9y) {};
\vertex[below=0.95cm of b9y] (b9yp) {};
\node[left=0.42cm of b9yp,dot,scale=1.8,white] (cdot) {};
\node[left=0.42cm of b9yp,thick,crossed dot] (cdot) {};
\end{feynman}
\begin{feynman}
\vertex (a);
\vertex[below=2.1cm of a] (b);
\vertex[right=0.89cm of b] (c) {(i)};
\vertex[right=3.24cm of c] (d) {(ii)};
\vertex[right=3.24cm of d] (e) {(iii)};
\vertex[right=3.24cm of e] (f) {(iv)};
\vertex[right=3.24cm of f] (f2) {(v)};
\vertex[below=3cm of a] (a2);
\vertex[below=2cm of a2] (b);
\vertex[right=0.89cm of b] (c) {(vi)};
\vertex[right=3.24cm of c] (d) {(vii)};
\vertex[right=3.24cm of d] (e) {(viii)};
\vertex[right=3.24cm of e] (f) {(ix)};
\end{feynman}
\end{tikzpicture}
\caption{Diagrams of topology (a) entering the $\mathcal{O}(G^2)$ sector of the 2PN dynamics. Henceforth, only powers of $v$ beyond $v^0$ are explicitly displayed in the vertices.}
\label{DiagsaG2order2PN}
\end{figure*}
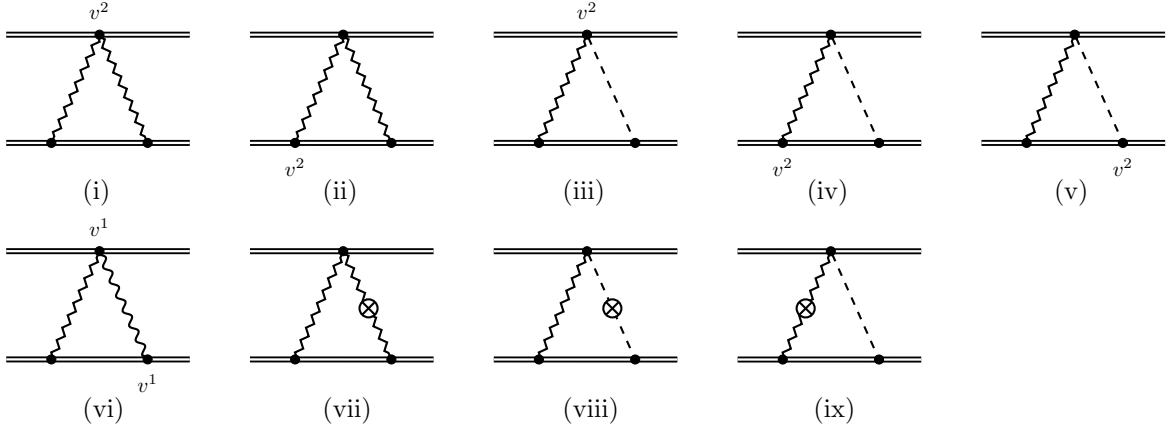
\begin{widetext}
\begin{align}
L_{\rm (i)} &= \frac{G^2 m^0_1 (m^0_2)^2 (\alpha_2^0)^2 ((\alpha_1^0)^2+\beta_1^0)}{4r^2} \V_1^2  + (1\leftrightarrow 2)\,, \\
L_{\rm (ii)} &= \frac{G^2 m^0_1 (m^0_2)^2 (\alpha_2^0)^2 ((\alpha_1^0)^2+\beta_1^0)}{2r^2} \V_2^2 + (1\leftrightarrow 2)\,, \\
L_{\rm (iii)} &= -\frac{3G^2 m^0_1 (m^0_2)^2 \alpha_1^0 \alpha_2^0}{2r^2} \V_1^2 + (1\leftrightarrow 2)\,, \\
L_{\rm (iv)} &= -\frac{3G^2 m^0_1 (m^0_2)^2 \alpha_1^0 \alpha_2^0}{2r^2} \V_2^2 + (1\leftrightarrow 2)\,, \\
L_{\rm (v)} &= \frac{G^2 m^0_1 (m^0_2)^2 \alpha_1^0 \alpha_2^0}{2r^2} \V_2^2 + (1\leftrightarrow 2)\,, \\
L_{\rm (vi)} &= \frac{4G^2 m^0_1 (m^0_2)^2 \alpha_1^0 \alpha_2^0}{r^2} \V_1\cdot\V_2 + (1\leftrightarrow 2) \,,\\
L_{\rm (vii)} &= \frac{G^2 m^0_1 (m^0_2)^2 (\alpha_2^0)^2 ((\alpha_1^0)^2 + \beta_1^0)}{2r} \left\{ \RR\cdot\A_2 - \frac{1}{r} [\V_2^2 - (\RR\cdot\V_2)^2] \right\} + (1\leftrightarrow 2) \,,\\
L_{\rm (viii)} &= \frac{G^2 m^0_1 (m^0_2)^2 \alpha_1^0 \alpha_2^0 }{2r} \left\{ \RR\cdot\A_2 - \frac{1}{r} [\V_2^2 - (\RR\cdot\V_2)^2] \right\} + (1\leftrightarrow 2) \,,\\
L_{\rm (ix)} &= \frac{G^2 m^0_1 (m^0_2)^2 \alpha_1^0 \alpha_2^0 }{2r} \left\{ \RR\cdot\A_2 - \frac{1}{r} [\V_2^2 - (\RR\cdot\V_2)^2] \right\} + (1\leftrightarrow 2)\,.
\end{align}
\end{widetext}

For the topology (b), we now have six diagrams contributing to the 2PN order. They are shown in Fig.~\ref{fig:diags1png200}.
These diagrams involve the four types of propagator present in our EFT: the one for the scalar field $\Phi$, as well as the three ones for the gravitational field, $\phi$, $A_i$, and $\sigma_{ij}$, the latter being represented by a thick wavy line. In these diagrams, the three-scalar-scalar-graviton vertices are derived from the minimal coupling of the scalar field to gravity, Eq.~\ref{phiaction0}, and read as 
\begin{widetext}
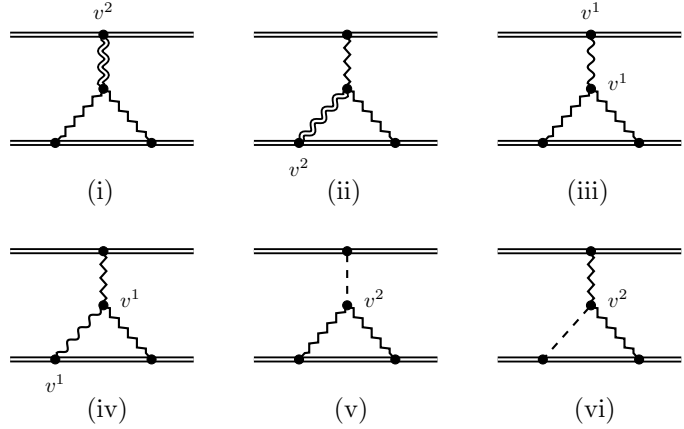
\begin{figure*}
\centering
\begin{tikzpicture}
\begin{feynman}[scale=0.8, transform shape]
\vertex (a);
\vertex[small,dot,right=1.5cm of a] (b) {};
\vertex[small,dot,below=0.9cm of b] (g) {};
\vertex[below=0.45cm of b] (g0) {};
\vertex[right=1.5cm of g0] (g00);
\vertex[right=1.5cm of b] (c);
\vertex[below=1.8cm of a] (d);
\vertex[right=1.5cm of d] (e);
\vertex[small,dot,right=0.7cm of d] (e00) {};
\vertex[small,dot,right=2.3cm of d] (e000) {};
\vertex[right=1.55cm of e] (f);
\vertex[below=2.5cm of b] (y) {};
\vertex[right=1cm of c] (a2);
\vertex[small,dot,right=1.5cm of a2] (b2) {};
\vertex[small,dot,below=0.9cm of b2] (g2) {};
\vertex[below=0.45cm of b2] (g20) {};
\vertex[right=1.5cm of g20] (g200);
\vertex[right=1.5cm of b2] (c2);
\vertex[below=1.8cm of a2] (d2);
\vertex[right=1.5cm of d2] (e2);
\vertex[small,dot,right=0.7cm of d2] (e200) {};
\vertex[small,dot,right=2.3cm of d2] (e2000) {};
\vertex[right=1.55cm of e2] (f2);
\vertex[below=2.5cm of b2] (y2) {};
\vertex[right=1cm of c2] (a3);
\vertex[small,dot,right=1.5cm of a3] (b3) {};
\vertex[small,dot,below=0.9cm of b3] (g3) {};
\vertex[below=0.45cm of b3] (g30) {};
\vertex[right=1.5cm of g30] (g300);
\vertex[right=1.5cm of b3] (c3);
\vertex[below=1.8cm of a3] (d3);
\vertex[right=1.5cm of d3] (e3);
\vertex[small,dot,right=0.7cm of d3] (e300) {};
\vertex[small,dot,right=2.3cm of d3] (e3000) {};
\vertex[right=1.55cm of e3] (f3);
\vertex[below=2.5cm of b3] (y3) {};
\vertex[below=1.8cm of d] (a4);
\vertex[small,dot,right=1.5cm of a4] (b4) {};
\vertex[small,dot,below=0.9cm of b4] (g4) {};
\vertex[below=0.45cm of b4] (g40) {};
\vertex[right=1.5cm of g40] (g400);
\vertex[right=1.5cm of b4] (c4);
\vertex[below=1.8cm of a4] (d4);
\vertex[right=1.5cm of d4] (e4);
\vertex[small,dot,right=0.7cm of d4] (e400) {};
\vertex[small,dot,right=2.3cm of d4] (e4000) {};
\vertex[right=1.55cm of e4] (f4);
\vertex[below=2.5cm of b4] (y4) {};
\vertex[right=1cm of c4] (a5);
\vertex[small,dot,right=1.5cm of a5] (b5) {};
\vertex[small,dot,below=0.9cm of b5] (g5) {};
\vertex[below=0.45cm of b5] (g50) {};
\vertex[right=1.5cm of g50] (g500);
\vertex[right=1.5cm of b5] (c5);
\vertex[below=1.8cm of a5] (d5);
\vertex[right=1.5cm of d5] (e5);
\vertex[small,dot,right=0.7cm of d5] (e500) {};
\vertex[small,dot,right=2.3cm of d5] (e5000) {};
\vertex[right=1.55cm of e5] (f5);
\vertex[below=2.5cm of b5] (y5) {};
\vertex[right=1cm of c5] (a6);
\vertex[small,dot,right=1.5cm of a6] (b6) {};
\vertex[small,dot,below=0.9cm of b6] (g6) {};
\vertex[below=0.45cm of b6] (g60) {};
\vertex[right=1.5cm of g60] (g600);
\vertex[right=1.5cm of b6] (c6);
\vertex[below=1.8cm of a6] (d6);
\vertex[right=1.5cm of d6] (e6);
\vertex[small,dot,right=0.7cm of d6] (e600) {};
\vertex[small,dot,right=2.3cm of d6] (e6000) {};
\vertex[right=1.55cm of e6] (f6);
\vertex[below=2.5cm of b6] (y6) {};
\diagram*{
(a) --[line width=0.8mm,plain] (c),
(a) --[line width=0.3mm,plain,white] (c),
(d) --[line width=0.8mm,plain] (f),
(d) --[line width=0.3mm,plain,white] (f),
(b) -- [\gravS] (g),
(b) -- [\gravSS] (g),
(g) -- [\scalar] (e00),
(g) -- [\scalar] (e000),
(a2) --[line width=0.8mm,plain] (c2),
(a2) --[line width=0.3mm,plain,white] (c2),
(d2) --[line width=0.8mm,plain] (f2),
(d2) --[line width=0.3mm,plain,white] (f2),
(b2) -- [\scalar] (g2),
(e200) -- [\gravS] (g2),
(e200) -- [\gravSS] (g2),
(g2) -- [\scalar] (e2000),
(a3) --[line width=0.8mm,plain] (c3),
(a3) --[line width=0.3mm,plain,white] (c3),
(d3) --[line width=0.8mm,plain] (f3),
(d3) --[line width=0.3mm,plain,white] (f3),
(b3) -- [\gravA] (g3),
(g3) -- [\scalar] (e300),
(g3) -- [\scalar] (e3000),
(a4) --[line width=0.8mm,plain] (c4),
(a4) --[line width=0.3mm,plain,white] (c4),
(d4) --[line width=0.8mm,plain] (f4),
(d4) --[line width=0.3mm,plain,white] (f4),
(b4) -- [\scalar] (g4),
(g4) -- [\gravA] (e400),
(g4) -- [\scalar] (e4000),
(a5) --[line width=0.8mm,plain] (c5),
(a5) --[line width=0.3mm,plain,white] (c5),
(d5) --[line width=0.8mm,plain] (f5),
(d5) --[line width=0.3mm,plain,white] (f5),
(b5) -- [\gravphi] (g5),
(g5) -- [\scalar] (e500),
(g5) -- [\scalar] (e5000),
(a6) --[line width=0.8mm,plain] (c6),
(a6) --[line width=0.3mm,plain,white] (c6),
(d6) --[line width=0.8mm,plain] (f6),
(d6) --[line width=0.3mm,plain,white] (f6),
(b6) -- [\scalar] (g6),
(g6) -- [\gravphi] (e600),
(g6) -- [\scalar] (e6000),
};
\vertex[dot,right=1.475cm of a] (bx) {};
\vertex[dot,right=0.675cm of d] (e00) {};
\vertex[dot,right=2.275cm of d] (e000) {};
\vertex[dot,below=0.9cm of bx] (g) {};
\vertex[above=0.4cm of bx] (v2) {$v^2$};
\vertex[below=0.4cm of e00] (v2) {};
\vertex[below=0.4cm of e000] (v2) {};
\vertex[right=0.45cm of g] (z) {};
\vertex[above=0.12cm of z] (v) {};
\vertex[dot,right=1.475cm of a2] (b2) {};
\vertex[dot,right=0.675cm of d2] (e200) {};
\vertex[dot,right=2.275cm of d2] (e2000) {};
\vertex[dot,below=0.9cm of b2] (g2) {};
\vertex[above=0.4cm of b2] (v2) {};
\vertex[below=0.4cm of e200] (v2) {$v^2$};
\vertex[below=0.4cm of e2000] (v2) {};
\vertex[right=0.45cm of g2] (z2) {};
\vertex[above=0.12cm of z2] (v2) {};
\vertex[dot,right=1.475cm of a3] (b3) {};
\vertex[dot,right=0.675cm of d3] (e300) {};
\vertex[dot,right=2.275cm of d3] (e3000) {};
\vertex[dot,below=0.9cm of b3] (g3) {};
\vertex[above=0.4cm of b3] (v3) {$v^1$};
\vertex[below=0.4cm of e300] (v3) {};
\vertex[below=0.4cm of e3000] (v3) {};
\vertex[right=0.45cm of g3] (z3) {};
\vertex[above=0.12cm of z3] (v3) {$v^1$};
\vertex[dot,right=1.475cm of a4] (b4) {};
\vertex[dot,right=0.675cm of d4] (e400) {};
\vertex[dot,right=2.275cm of d4] (e4000) {};
\vertex[dot,below=0.9cm of b4] (g4) {};
\vertex[above=0.4cm of b4] (v4) {};
\vertex[below=0.4cm of e400] (v4) {$v^1$};
\vertex[below=0.4cm of e4000] (v4) {};
\vertex[right=0.45cm of g4] (z4) {};
\vertex[above=0.12cm of z4] (v4) {$v^1$};
\vertex[dot,right=1.475cm of a5] (b5) {};
\vertex[dot,right=0.675cm of d5] (e500) {};
\vertex[dot,right=2.275cm of d5] (e5000) {};
\vertex[dot,below=0.9cm of b5] (g5) {};
\vertex[above=0.4cm of b5] (v5) {};
\vertex[below=0.4cm of e500] (v5) {};
\vertex[below=0.4cm of e5000] (v5) {};
\vertex[right=0.45cm of g5] (z5) {};
\vertex[above=0.12cm of z5] (v5) {$v^2$};
\vertex[dot,right=1.475cm of a6] (b6) {};
\vertex[dot,right=0.675cm of d6] (e600) {};
\vertex[dot,right=2.275cm of d6] (e6000) {};
\vertex[dot,below=0.9cm of b6] (g6) {};
\vertex[above=0.4cm of b6] (v6) {};
\vertex[below=0.4cm of e600] (v6) {};
\vertex[below=0.4cm of e6000] (v6) {};
\vertex[right=0.45cm of g6] (z6) {};
\vertex[above=0.12cm of z6] (v6) {$v^2$};
\end{feynman}
\begin{feynman}
\vertex (a);
\vertex[below=2.1cm of a] (b);
\vertex[right=0.89cm of b] (c) {(i)};
\vertex[right=3.24cm of c] (d) {(ii)};
\vertex[right=3.24cm of d] (e) {(iii)};
\vertex[below=3cm of a] (a2);
\vertex[below=2cm of a2] (b);
\vertex[right=0.89cm of b] (c) {(iv)};
\vertex[right=3.24cm of c] (d) {(v)};
\vertex[right=3.24cm of d] (e) {(vi)};
\end{feynman}
\end{tikzpicture}
\caption{Diagrams of topology (b) entering the $\mathcal{O}(G^2)$ sector of the 2PN dynamics. Thick wavy lines represent $\sigma$ mode propagators.}
\label{fig:diags1png200}
\end{figure*}
\begin{equation}
S_{\vphi} \rightarrow -\frac{4}{\Lambda} \int d^{d+1}x\,\left[ \left( c_d \phi -\frac12 \sigma \right) (\partial_0\Phi)^2 + 2 A_i \partial_0\Phi \partial_i\Phi
+ \left( \frac12 \sigma \delta_{ij} - \sigma_{ij} \right) \partial_i\Phi \partial_j\Phi 
\right]\,.
\end{equation}
\end{widetext}

In particular, for the computation of these diagrams, the following one-loop master integral is needed (along with its generalization to include up to two factors of $q_i$s in the numerator, obtained from the same expression through tensor reduction):
\begin{widetext}
\begin{equation}\label{oneloopmasterintss}
\int_\Q \frac{1}{(\Q^2)^\alpha [(\K+\Q)^2]^\beta} = \frac{1}{(4\pi)^{d/2}} \frac{\Gamma(\alpha+\beta-d/2)}{\Gamma(\alpha)\Gamma(\beta)} \frac{\Gamma(d/2-\alpha)\Gamma(d/2-\beta)}{\Gamma(d-\alpha-\beta)} (\K^2)^{d/2-\alpha-\beta}\,.
\end{equation}
\end{widetext}

Computation of diagrams (i)-(vi) results in
\begin{widetext}
\begin{align}
L_{\rm (i)} &= -\frac{G^2 m^0_1 (m^0_2)^2 (\alpha^0_2)^2}{2r^2} [\V_1^2 - (\RR\cdot\V_1)^2] + (1\leftrightarrow 2) \,,\\
L_{\rm (ii)} &= \frac{2G^2 m^0_1 (m^0_2)^2 \alpha^0_1 \alpha^0_2}{r^2} [\V_2^2 - 2(\RR\cdot\V_2)^2] +  (1\leftrightarrow 2) \,, \\
L_{\rm (iii)} &= \frac{G^2 m^0_1 (m^0_2)^2 (\alpha^0_2)^2}{r^2} [\V_1\cdot\V_2 - (\RR\cdot\V_1)(\RR\cdot\V_2)]  + (1\leftrightarrow 2) \,,\\
L_{\rm (iv)} &= -\frac{2G^2 m^0_1 (m^0_2)^2 \alpha^0_1 \alpha^0_2}{r^2} [\V_2^2 + \V_1\cdot\V_2 - 2 (\RR\cdot\V_2) (\V_1+\V_2)\cdot\RR ]  + (1\leftrightarrow 2)\,, \\
L_{\rm (v)} &= - \frac{G^2 m^0_1 (m^0_2)^2 (\alpha^0_2)^2}{2r^2} [\V_2^2 - (\RR\cdot\V_2)^2]  + (1\leftrightarrow 2)\,,\\
L_{\rm (vi)} &= \frac{2G^2 m^0_1 (m^0_2)^2 \alpha^0_1 \alpha^0_2}{r^2} [\V_1\cdot\V_2 - 2 (\RR\cdot\V_1)(\RR\cdot\V_2)]  + (1\leftrightarrow 2)\,.
\end{align}
\end{widetext}

\subsection{$\mathcal{O}(G^3)$ contributions}

All the diagram topologies of $\mathcal{O}(G^3)$ displayed in Fig.~\ref{fig:topologyG3} present two-loop integrals. 
Since these diagrams already give contributions at order $G^2$ beyond the Newtonian dynamics, neither velocity corrections in the vertices, propagator insertions, nor time derivatives, contribute to the 2PN order. Because of this, the only relevant gravitational worldline couplings in these static diagrams is the one involving the $\phi$ field. 

The properties above allow us to reduce the number of topologies from nine to five as we explain next. 
In topologies (c) and (e), only the part involving spatial derivatives in the four vertex $\phi\phi\Phi\Phi$ derived from Eq.~\eqref{phiaction0} would contribute. However, in the Kaluza-Klein decomposition, it is easy to check that $\sqrt{-g}g^{ij} =0$, when the modes $\sigma_{ij}$ and $A_i$ are set to zero. Similarly, because both three vertices $\phi\phi\phi$ and $\phi\Phi\Phi$ present time derivatives, diagrams (d) and (f) give automatically vanishing contributions to the 2PN order. Thus, we immediately have
\begin{equation}
L_{(c)} = L_{(d)} = L_{(e)} = L_{(f)} = 0\,.
\end{equation}

For the other diagrams, we have the general structure:
\begin{equation}
\int_\K e^{i \K\cdot\R} \times \text{(two-loop integral)}\,.
\end{equation}
In particular, the two-loop integrals present in the diagrams with topologies (a), (b), (g), and (i) can all be reduced (by performing momentum shifts) to nested one-loop integrals [for diagrams (a), (b), and (g)], or can be factorizable to products of two one-loop integrals [in the case of diagram (i)], with the same form of the family of one-loop integrals given by Eq.~\eqref{oneloopmasterintss}. Then, Eq.~\eqref{intKmaster} is used to perform the integration of the remaining expressions over $\int_\K e^{i\K\cdot\R}$.   

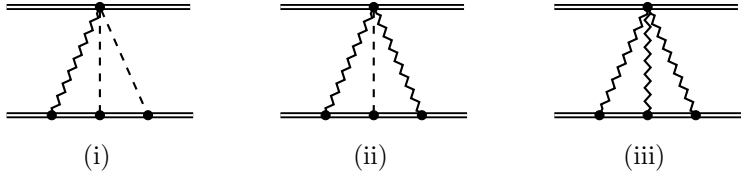
\begin{figure*}
\centering
\begin{tikzpicture}
\begin{feynman}[scale=0.8, transform shape]
\vertex (a);
\vertex[small,dot,right=1.5cm of a] (b) {};
\vertex[small,dot,right=0.7cm of a] (b0);
\vertex[small,dot,right=2.3cm of a] (b00);
\vertex[right=1.5cm of b] (c);
\vertex[below=1.8cm of a] (d);
\vertex[small,dot,right=1.5cm of d] (e) {};
\vertex[small,dot,right=0.7cm of d] (e0) {};
\vertex[small,dot,right=2.3cm of d] (e00) {};
\vertex[right=1.55cm of e] (f);
\vertex[right=1.5cm of c] (a2);
\vertex[small,dot,right=1.5cm of a2] (b2) {};
\vertex[small,dot,right=0.7cm of a2] (b20);
\vertex[small,dot,right=2.3cm of a2] (b200);
\vertex[right=1.5cm of b2] (c2);
\vertex[below=1.8cm of a2] (d2);
\vertex[small,dot,right=1.5cm of d2] (e2) {};
\vertex[small,dot,right=0.7cm of d2] (e20) {};
\vertex[small,dot,right=2.3cm of d2] (e200) {};
\vertex[right=1.55cm of e2] (f2);
\vertex[right=1.5cm of c2] (a3);
\vertex[small,dot,right=1.5cm of a3] (b3) {};
\vertex[small,dot,right=0.7cm of a3] (b30);
\vertex[small,dot,right=2.3cm of a3] (b300);
\vertex[right=1.5cm of b3] (c3);
\vertex[below=1.8cm of a3] (d3);
\vertex[small,dot,right=1.5cm of d3] (e3) {};
\vertex[small,dot,right=0.7cm of d3] (e30) {};
\vertex[small,dot,right=2.3cm of d3] (e300) {};
\vertex[right=1.55cm of e3] (f3);
\diagram*{
(a) --[line width=0.8mm,plain] (c),
(a) --[line width=0.3mm,plain,white] (c),
(d) --[line width=0.8mm,plain] (f),
(d) --[line width=0.3mm,plain,white] (f),
(e0) -- [\scalar] (b),
(b) -- [\gravphi] (e00),
(b) -- [\gravphi] (e),
(a2) --[line width=0.8mm,plain] (c2),
(a2) --[line width=0.3mm,plain,white] (c2),
(d2) --[line width=0.8mm,plain] (f2),
(d2) --[line width=0.3mm,plain,white] (f2),
(e20) -- [\scalar] (b2),
(e200) -- [\scalar] (b2),
(b2) -- [\gravphi] (e2),
(a3) --[line width=0.8mm,plain] (c3),
(a3) --[line width=0.3mm,plain,white] (c3),
(d3) --[line width=0.8mm,plain] (f3),
(d3) --[line width=0.3mm,plain,white] (f3),
(b3) -- [\scalar] (e30),
(e300) -- [\scalar] (b3),
(b3) -- [\scalar] (e3),
};
\vertex[dot,right=1.475cm of a] (bx) {};
\vertex[dot,right=1.475cm of d] (e) {};
\vertex[dot,right=0.675cm of d] (e0) {};
\vertex[dot,right=2.275cm of d] (e00) {};
\vertex[dot,right=1.475cm of a2] (bx) {};
\vertex[dot,right=1.475cm of d2] (e) {};
\vertex[dot,right=0.675cm of d2] (e0) {};
\vertex[dot,right=2.275cm of d2] (e00) {};
\vertex[dot,right=1.475cm of a3] (bx) {};
\vertex[dot,right=1.475cm of d3] (e) {};
\vertex[dot,right=0.675cm of d3] (e0) {};
\vertex[dot,right=2.275cm of d3] (e00) {};
\end{feynman}
\begin{feynman}
\vertex (a);
\vertex[below=2cm of a] (b);
\vertex[right=0.89cm of b] (c) {(i)};
\vertex[right=3.64cm of c] (d) {(ii)};
\vertex[right=3.64cm of d] (e) {(iii)};
\end{feynman}
\end{tikzpicture}
\caption{Diagrams of topology (a) entering the $\mathcal{O}(G^3)$ sector of the 2PN dynamics.}
\label{OG3diags2PNa}
\end{figure*}

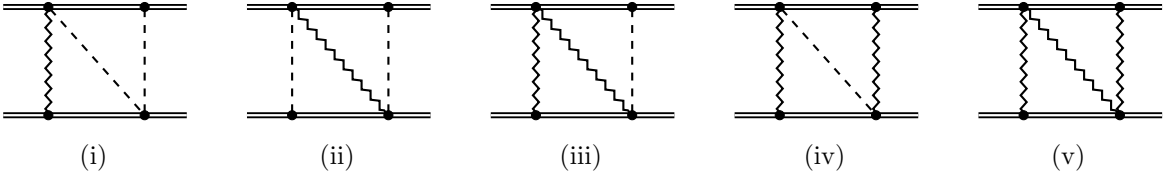
\begin{figure*}
\centering
\begin{tikzpicture}
\begin{feynman}[scale=0.8, transform shape]
\vertex(a);
\vertex[small,dot,white,right=1.5cm of a] (b) {};
\vertex[small,dot,right=0.7cm of a] (b0) {};
\vertex[small,dot,right=2.3cm of a] (b00) {};
\vertex[right=1.5cm of b] (c);
\vertex[below=1.8cm of a] (d);
\vertex[small,dot,right=1.5cm of d] (e);
\vertex[small,dot,right=0.7cm of d] (e0) {};
\vertex[small,dot,right=2.3cm of d] (e00) {};
\vertex[right=1.55cm of e] (f);
\vertex[dot,below=0.9cm of b] (g);
\vertex[right=1cm of c] (a2);
\vertex[small,dot,white,right=1.5cm of a2] (b2) {};
\vertex[small,dot,right=0.7cm of a2] (b20) {};
\vertex[small,dot,right=2.3cm of a2] (b200) {};
\vertex[right=1.5cm of b2] (c2);
\vertex[below=1.8cm of a2] (d2);
\vertex[small,dot,right=1.5cm of d2] (e2);
\vertex[small,dot,right=0.7cm of d2] (e20) {};
\vertex[small,dot,right=2.3cm of d2] (e200) {};
\vertex[right=1.55cm of e2] (f2);
\vertex[dot,below=0.9cm of b2] (g2);
\vertex[right=1cm of c2] (a3);
\vertex[small,dot,white,right=1.5cm of a3] (b3) {};
\vertex[small,dot,right=0.7cm of a3] (b30) {};
\vertex[small,dot,right=2.3cm of a3] (b300) {};
\vertex[right=1.5cm of b3] (c3);
\vertex[below=1.8cm of a3] (d3);
\vertex[small,dot,right=1.5cm of d3] (e3);
\vertex[small,dot,right=0.7cm of d3] (e30) {};
\vertex[small,dot,right=2.3cm of d3] (e300) {};
\vertex[right=1.55cm of e3] (f3);
\vertex[dot,below=0.9cm of b3] (g3);
\vertex[right=1cm of c3] (a4);
\vertex[small,dot,white,right=1.5cm of a4] (b4) {};
\vertex[small,dot,right=0.7cm of a4] (b40) {};
\vertex[small,dot,right=2.3cm of a4] (b400) {};
\vertex[right=1.5cm of b4] (c4);
\vertex[below=1.8cm of a4] (d4);
\vertex[small,dot,right=1.5cm of d4] (e4);
\vertex[small,dot,right=0.7cm of d4] (e40) {};
\vertex[small,dot,right=2.3cm of d4] (e400) {};
\vertex[right=1.55cm of e4] (f4);
\vertex[dot,below=0.9cm of b4] (g4);
\vertex[right=1cm of c4] (a5);
\vertex[small,dot,white,right=1.5cm of a5] (b5) {};
\vertex[small,dot,right=0.7cm of a5] (b50) {};
\vertex[small,dot,right=2.3cm of a5] (b500) {};
\vertex[right=1.5cm of b5] (c5);
\vertex[below=1.8cm of a5] (d5);
\vertex[small,dot,right=1.5cm of d5] (e5);
\vertex[small,dot,right=0.7cm of d5] (e50) {};
\vertex[small,dot,right=2.3cm of d5] (e500) {};
\vertex[right=1.55cm of e5] (f5);
\vertex[dot,below=0.9cm of b5] (g5);
\diagram*{
(a) --[line width=0.8mm,plain] (c),
(a) --[line width=0.3mm,plain,white] (c),
(d) --[line width=0.8mm,plain] (f),
(d) --[line width=0.3mm,plain,white] (f),
(e0) -- [\scalar] (b0),
(b0) -- [\gravphi] (e00),
(e00) -- [\gravphi] (b00),
(a2) --[line width=0.8mm,plain] (c2),
(a2) --[line width=0.3mm,plain,white] (c2),
(d2) --[line width=0.8mm,plain] (f2),
(d2) --[line width=0.3mm,plain,white] (f2),
(e20) -- [\gravphi] (b20),
(b20) -- [\scalar] (e200),
(e200) -- [\gravphi] (b200),
(a3) --[line width=0.8mm,plain] (c3),
(a3) --[line width=0.3mm,plain,white] (c3),
(d3) --[line width=0.8mm,plain] (f3),
(d3) --[line width=0.3mm,plain,white] (f3),
(e30) -- [\scalar] (b30),
(b30) -- [\scalar] (e300),
(e300) -- [\gravphi] (b300),
(a4) --[line width=0.8mm,plain] (c4),
(a4) --[line width=0.3mm,plain,white] (c4),
(d4) --[line width=0.8mm,plain] (f4),
(d4) --[line width=0.3mm,plain,white] (f4),
(e40) -- [\scalar] (b40),
(b40) -- [\gravphi] (e400),
(e400) -- [\scalar] (b400),
(a5) --[line width=0.8mm,plain] (c5),
(a5) --[line width=0.3mm,plain,white] (c5),
(d5) --[line width=0.8mm,plain] (f5),
(d5) --[line width=0.3mm,plain,white] (f5),
(e50) -- [\scalar] (b50),
(b50) -- [\scalar] (e500),
(e500) -- [\scalar] (b500),
};
\vertex[dot,right=0.675cm of a] (b0) {};
\vertex[dot,right=2.275cm of a] (b00) {};
\vertex[dot,right=0.675cm of d] (e0) {};
\vertex[dot,right=2.275cm of d] (e00) {};
\vertex[dot,right=0.675cm of a2] (b20) {};
\vertex[dot,right=2.275cm of a2] (b200) {};
\vertex[dot,right=0.675cm of d2] (e20) {};
\vertex[dot,right=2.275cm of d2] (e200) {};
\vertex[dot,right=0.675cm of a3] (b30) {};
\vertex[dot,right=2.275cm of a3] (b300) {};
\vertex[dot,right=0.675cm of d3] (e30) {};
\vertex[dot,right=2.275cm of d3] (e300) {};
\vertex[dot,right=0.675cm of a4] (b40) {};
\vertex[dot,right=2.275cm of a4] (b400) {};
\vertex[dot,right=0.675cm of d4] (e40) {};
\vertex[dot,right=2.275cm of d4] (e400) {};
\vertex[dot,right=0.675cm of a5] (b50) {};
\vertex[dot,right=2.275cm of a5] (b500) {};
\vertex[dot,right=0.675cm of d5] (e50) {};
\vertex[dot,right=2.275cm of d5] (e500) {};
\end{feynman}
\begin{feynman}
\vertex (a);
\vertex[below=2cm of a] (b);
\vertex[right=0.89cm of b] (c) {(i)};
\vertex[right=3.24cm of c] (d) {(ii)};
\vertex[right=3.24cm of d] (e) {(iii)};
\vertex[right=3.24cm of e] (f) {(iv)};
\vertex[right=3.24cm of f] (f2) {(v)};
\end{feynman}
\end{tikzpicture}
\caption{Diagrams of topology (b) entering the $\mathcal{O}(G^3)$ sector of the 2PN dynamics.}
\label{OG3diags2PNb}
\end{figure*}

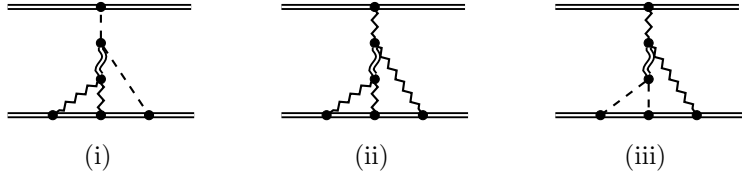
\begin{figure*}
\centering
\begin{tikzpicture}
\begin{feynman}[scale=0.8, transform shape]
\vertex (a);
\vertex[small,dot,right=1.5cm of a] (b) {};
\vertex[right=1.5cm of b] (c);
\vertex[below=1.8cm of a] (d);
\vertex[small,dot,right=1.5cm of d] (e) {};
\vertex[small,dot,right=0.7cm of d] (e0) {};
\vertex[small,dot,right=2.3cm of d] (e00) {};
\vertex[right=1.55cm of e] (f);
\vertex[dot,below=0.6cm of b] (h) {};
\vertex[dot,above=0.6cm of e] (h0) {};
\vertex[right=1.5cm of c] (a2);
\vertex[small,dot,right=1.5cm of a2] (b2) {};
\vertex[right=1.5cm of b2] (c2);
\vertex[below=1.8cm of a2] (d2);
\vertex[small,dot,right=1.5cm of d2] (e2) {};
\vertex[small,dot,right=0.7cm of d2] (e20) {};
\vertex[small,dot,right=2.3cm of d2] (e200) {};
\vertex[right=1.55cm of e2] (f2);
\vertex[dot,below=0.6cm of b2] (h2) {};
\vertex[dot,above=0.6cm of e2] (h20) {};
\vertex[right=1.5cm of c2] (a3);
\vertex[small,dot,right=1.5cm of a3] (b3) {};
\vertex[right=1.5cm of b3] (c3);
\vertex[below=1.8cm of a3] (d3);
\vertex[small,dot,right=1.5cm of d3] (e3) {};
\vertex[small,dot,right=0.7cm of d3] (e30) {};
\vertex[small,dot,right=2.3cm of d3] (e300) {};
\vertex[right=1.55cm of e3] (f3);
\vertex[dot,below=0.6cm of b3] (h3) {};
\vertex[dot,above=0.6cm of e3] (h30) {};
\diagram*{
(a) --[line width=0.8mm,plain] (c),
(a) --[line width=0.3mm,plain,white] (c),
(d) --[line width=0.8mm,plain] (f),
(d) --[line width=0.3mm,plain,white] (f),
(h) -- [\gravphi] (b),
(h) -- [\gravS] (h0),
(h) -- [\gravSS] (h0),
(h0) -- [\scalar] (e),
(h0) -- [\scalar] (e0),
(h) -- [\gravphi] (e00),
(a2) --[line width=0.8mm,plain] (c2),
(a2) --[line width=0.3mm,plain,white] (c2),
(d2) --[line width=0.8mm,plain] (f2),
(d2) --[line width=0.3mm,plain,white] (f2),
(b2) -- [\scalar] (h2),
(h2) -- [\gravS] (h20),
(h2) -- [\gravSS] (h20),
(h20) -- [\scalar] (e2),
(h20) -- [\scalar] (e20),
(h2) -- [\scalar] (e200),
(a3) --[line width=0.8mm,plain] (c3),
(a3) --[line width=0.3mm,plain,white] (c3),
(d3) --[line width=0.8mm,plain] (f3),
(d3) --[line width=0.3mm,plain,white] (f3),
(b3) -- [\scalar] (h3),
(h3) -- [\gravS] (h30),
(h3) -- [\gravSS] (h30),
(h30) -- [\gravphi] (e3),
(h30) -- [\gravphi] (e30),
(h3) -- [\scalar] (e300),
};
\vertex[dot,right=1.475cm of a] (b7x) {};
\vertex[dot,right=1.475cm of d] (e7) {};
\vertex[dot,right=0.675cm of d] (e70) {};
\vertex[dot,right=2.275cm of d] (e700) {};
\vertex[dot,right=1.475cm of a2] (b7x) {};
\vertex[dot,right=1.475cm of d2] (e7) {};
\vertex[dot,right=0.675cm of d2] (e70) {};
\vertex[dot,right=2.275cm of d2] (e700) {};
\vertex[dot,right=1.475cm of a3] (b7x) {};
\vertex[dot,right=1.475cm of d3] (e7) {};
\vertex[dot,right=0.675cm of d3] (e70) {};
\vertex[dot,right=2.275cm of d3] (e700) {};

\end{feynman}
\begin{feynman}
\vertex (a);
\vertex[below=2cm of a] (b);
\vertex[right=0.89cm of b] (c) {(i)};
\vertex[right=3.64cm of c] (d) {(ii)};
\vertex[right=3.64cm of d] (e) {(iii)};
\end{feynman}
\end{tikzpicture}
\caption{Diagrams of topology (g) entering the $\mathcal{O}(G^3)$ sector of the 2PN dynamics.}
\label{OG3diags2PNg}
\end{figure*}

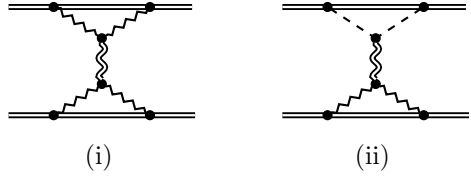
\begin{figure*}
\centering
\begin{tikzpicture}
\begin{feynman}[scale=0.8, transform shape]
\vertex (a8);
\vertex[small,dot,white,right=1.5cm of a8] (b8) {};
\vertex[small,dot,right=0.7cm of a8] (b80) {};
\vertex[small,dot,right=2.3cm of a8] (b800) {};
\vertex[right=1.5cm of b8] (c8);
\vertex[below=1.8cm of a8] (d8);
\vertex[small,dot,white,right=1.5cm of d8] (e8) {};
\vertex[small,dot,right=0.7cm of d8] (e80) {};
\vertex[small,dot,right=2.3cm of d8] (e800) {};
\vertex[right=1.55cm of e8] (f8);
\vertex[dot,below=0.52cm of b8] (h8) {};
\vertex[dot,above=0.52cm of e8] (h80) {};
\vertex[right=1.5cm of c8] (a9);
\vertex[small,dot,white,right=1.5cm of a9] (b9) {};
\vertex[small,dot,right=0.7cm of a9] (b90) {};
\vertex[small,dot,right=2.3cm of a9] (b900) {};
\vertex[right=1.5cm of b9] (c9);
\vertex[below=1.8cm of a9] (d9);
\vertex[small,dot,white,right=1.5cm of d9] (e9) {};
\vertex[small,dot,right=0.7cm of d9] (e90) {};
\vertex[small,dot,right=2.3cm of d9] (e900) {};
\vertex[right=1.55cm of e9] (f9);
\vertex[dot,below=0.52cm of b9] (h9) {};
\vertex[dot,above=0.52cm of e9] (h90) {};
\diagram*{
(a8) --[line width=0.8mm,plain] (c8),
(a8) --[line width=0.3mm,plain,white] (c8),
(d8) --[line width=0.8mm,plain] (f8),
(d8) --[line width=0.3mm,plain,white] (f8),
(b80) -- [\scalar] (h8),
(b800) -- [\scalar] (h8),
(h8) -- [\gravS] (h80),
(h8) -- [\gravSS] (h80),
(h80) -- [\scalar] (e80),
(h80) -- [\scalar] (e800),
(a9) --[line width=0.8mm,plain] (c9),
(a9) --[line width=0.3mm,plain,white] (c9),
(d9) --[line width=0.8mm,plain] (f9),
(d9) --[line width=0.3mm,plain,white] (f9),
(b90) -- [\gravphi] (h9),
(b900) -- [\gravphi] (h9),
(h9) -- [\gravS] (h90),
(h9) -- [\gravSS] (h90),
(h90) -- [\scalar] (e90),
(h90) -- [\scalar] (e900),
};
\vertex[dot,right=0.675cm of a8] (b80x) {};
\vertex[dot,right=2.275cm of a8] (b800) {};
\vertex[dot,right=0.675cm of d8] (e80) {};
\vertex[dot,right=2.275cm of d8] (e800) {};
\vertex[dot,right=0.675cm of a9] (b90x) {};
\vertex[dot,right=2.275cm of a9] (b900) {};
\vertex[dot,right=0.675cm of d9] (e90) {};
\vertex[dot,right=2.275cm of d9] (e900) {};
\end{feynman}
\begin{feynman}
\vertex (a);
\vertex[below=2cm of a] (b);
\vertex[right=0.89cm of b] (c) {(i)};
\vertex[right=3.64cm of c] (d) {(ii)};
\end{feynman}
\end{tikzpicture}
\caption{Diagrams of topology (i) entering the $\mathcal{O}(G^3)$ sector of the 2PN dynamics.}
\label{OG3diags2PNi}
\end{figure*}

All the individual diagrams contributing to the 2PN conservative binary dynamics from topologies (a), (b), (g), and (i) are given in Figs.~\ref{OG3diags2PNa}, \ref{OG3diags2PNb}, \ref{OG3diags2PNg}, and \ref{OG3diags2PNi}, respectively. Feynman rules are then employed and computation follows straightforwardly, without any complications. The results are presented below.

From topology (a), Fig.~\ref{OG3diags2PNa}:
\begin{align}
L_{\rm (i)}^{(a)} & = \frac{G^3 m^0_1 (m^0_2)^3 \alpha_1^0\alpha_2^0}{2r^3} + (1 \leftrightarrow 2) \,,\\
L_{\rm (ii)}^{(a)} & = \frac{G^3 m^0_1 (m^0_2)^3 (\alpha_2^0)^2 ((\alpha_1^0)^2+\beta_1^0)}{2r^3} + (1 \leftrightarrow 2) \,,\\
L_{\rm (iii)}^{(a)} & = \frac{G^3 m^0_1 (m^0_2)^3 (\alpha_2^0)^3 [\alpha_1^0((\alpha_1^0)^2+3\beta_1^0)+\beta_1^{'0}]}{6r^3} + (1 \leftrightarrow 2) \,.
\end{align}

From topology (b), Fig.~\ref{OG3diags2PNb}:
\begin{align}
L^{(b)}_{\rm (i)} & = \frac{G^3 (m^0_1)^2 (m^0_2)^2 \alpha_1^0\alpha_2^0}{r^3} + (1 \leftrightarrow 2)  \,,\\
L^{(b)}_{\rm (ii)} & = \frac{G^3 (m^0_1)^2 (m^0_2)^2 \alpha_1^0\alpha_2^0}{r^3}  \,,\\
L^{(b)}_{\rm (iii)} & = \frac{G^3 (m^0_1)^2 (m^0_2)^2 (\alpha_1^0)^2 ((\alpha_2^0)^2+\beta_2^0)}{r^3} + (1 \leftrightarrow 2) \,,\\
L^{(b)}_{\rm (iv)} & = \frac{G^3 (m^0_1)^2 (m^0_2)^2 (\alpha_1^0)^2 (\alpha_2^0)^2}{r^3}  \,,\\
L^{(b)}_{\rm (v)} & = \frac{G^3 (m^0_1)^2 (m^0_2)^2 \alpha_1^0 \alpha_2^0 [(\alpha_1^0)^2+\beta_1^0] [(\alpha_2^0)^2+\beta_2^0] }{r^3} \,.
\end{align}

From topology (g), Fig.~\ref{OG3diags2PNg}:
\begin{align}
L^{(g)}_{\rm (i)} & = \frac{G^3 m^0_1 (m^0_2)^3 (\alpha_2^0)^2}{3r^3} + (1 \leftrightarrow 2)\,,\\
L^{(g)}_{\rm (ii)} & = \frac{G^3 m^0_1 (m^0_2)^3 \alpha_1 (\alpha_2^0)^3}{3r^3} + (1 \leftrightarrow 2) \,.\\
L^{(g)}_{\rm (iii)} & = \frac{G^3 m^0_1 (m^0_2)^3 \alpha_1^0 \alpha_2^0}{3r^3} + (1 \leftrightarrow 2) \,,
\end{align}

From topology (i), Fig.~\ref{OG3diags2PNi}:
\begin{equation}
L^{(i)}_{\rm (i)} = L^{(i)}_{\rm (ii)}  = 0 \,.
\end{equation} 
As in the GR case, this topology yields vanishing results in the sense that $L\sim \delta(r)$, thus not being relevant in the near description of the binary dynamics \cite{Gilmore:2008gq}.

Finally, for the contributions from diagrams with topology (h), we need to compute the following two-loop integral with momenta in its numerator:
\begin{equation}\label{difintegral}
\int_{\PP\Q} \frac{p^i q^j (k-p)^k (k+q)^l}{(\K-\PP)^2 (\K+\Q)^2 (\PP+\Q)^2 \PP^2 \Q^2}\,.
\end{equation}
Tensor reduction can be used in this case to obtain a generic expression for this integral in terms of scalar two-loop integrals. In particular, most of the resulting integrals are either factorizable or nested, given in terms of one-loop integrals with the form of Eq.~\eqref{oneloopmasterintss}. There are also appearances of the scalar two-loop with same denominator of Eq.~\eqref{difintegral}, which, in turn, can be reduced to one-loop integrals by using the integration by parts technique \cite{Smirnov:2004ym}. In this case, one can derive \cite{Gilmore:2008gq}
\begin{align}
&\int_{\PP\Q} \frac{1}{(\K-\PP)^2 (\K+\Q)^2 (\PP+\Q)^2 \PP^2 \Q^2}    \\
&\qquad\qquad= \frac{2}{d-4} \int_{\PP\Q} \bigg[ \frac{1}{\PP^2\Q^2 (\K+\Q)^2 (\K+\PP)^4} - \frac{1}{\PP^2\Q^2 (\PP+\Q)^2 (\K+\PP)^4} \bigg]\,.
\end{align}
Then, we notice that the first integral on the RHS of this expression is factorizable into a product of two one-loop integrals, while the second one is nested, and easily computed by first performing the integration over $\Q$, followed by integration over $\PP$. In all these cases, the one-loop integral \eqref{oneloopmasterintss} is used. 
\begin{figure*}
\centering
\begin{tikzpicture}
\begin{feynman}[scale=0.8, transform shape]
\vertex (a8);
\vertex[small,dot,white,right=1.5cm of a8] (b8) {};
\vertex[small,dot,right=0.7cm of a8] (b80) {};
\vertex[small,dot,right=2.3cm of a8] (b800) {};
\vertex[right=1.5cm of b8] (c8);
\vertex[below=1.8cm of a8] (d8);
\vertex[small,dot,right=1.5cm of d8] (e8);
\vertex[small,dot,right=0.7cm of d8] (e80) {};
\vertex[small,dot,right=2.3cm of d8] (e800) {};
\vertex[right=1.55cm of e8] (f8);
\vertex[dot,below=0.9cm of b80] (g80) {};
\vertex[dot,below=0.9cm of b800] (g800) {};
\vertex[right=1.5cm of c8] (a9);
\vertex[small,dot,white,right=1.5cm of a9] (b9) {};
\vertex[small,dot,right=0.7cm of a9] (b90) {};
\vertex[small,dot,right=2.3cm of a9] (b900) {};
\vertex[right=1.5cm of b9] (c9);
\vertex[below=1.8cm of a9] (d9);
\vertex[small,dot,right=1.5cm of d9] (e9);
\vertex[small,dot,right=0.7cm of d9] (e90) {};
\vertex[small,dot,right=2.3cm of d9] (e900) {};
\vertex[right=1.55cm of e9] (f9);
\vertex[dot,below=0.9cm of b90] (g90) {};
\vertex[dot,below=0.9cm of b900] (g900) {};
\diagram*{
(a8) --[line width=0.8mm,plain] (c8),
(a8) --[line width=0.3mm,plain,white] (c8),
(d8) --[line width=0.8mm,plain] (f8),
(d8) --[line width=0.3mm,plain,white] (f8),
(b80) -- [\scalar] (e80),
(b800) -- [\gravphi] (e800),
(g80) -- [\gravS] (g800),
(g80) -- [\gravSS] (g800),
(a9) --[line width=0.8mm,plain] (c9),
(a9) --[line width=0.3mm,plain,white] (c9),
(d9) --[line width=0.8mm,plain] (f9),
(d9) --[line width=0.3mm,plain,white] (f9),
(b90) -- [\scalar] (e90),
(b900) -- [\scalar] (e900),
(g90) -- [\gravS] (g900),
(g90) -- [\gravSS] (g900),
};
\vertex[dot,right=0.675cm of a8] (b80x) {};
\vertex[dot,right=2.275cm of a8] (b800) {};
\vertex[dot,right=0.675cm of d8] (e80) {};
\vertex[dot,right=2.275cm of d8] (e800) {};
\vertex[dot,right=0.675cm of a9] (b90x) {};
\vertex[dot,right=2.275cm of a9] (b900) {};
\vertex[dot,right=0.675cm of d9] (e90) {};
\vertex[dot,right=2.275cm of d9] (e900) {};
\end{feynman}
\begin{feynman}
\vertex (a);
\vertex[below=2cm of a] (b);
\vertex[right=0.89cm of b] (c) {(a)};
\vertex[right=3.64cm of c] (d) {(b)};
\end{feynman}
\end{tikzpicture}
\caption{Diagrams of topology (h) entering the $\mathcal{O}(G^3)$ sector of the 2PN dynamics.}
\label{OG3diags2PNh}
\end{figure*}
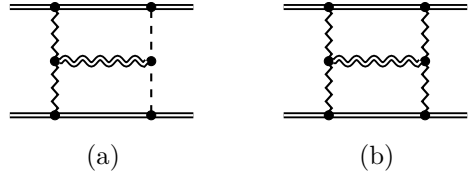

Thus, using the above results, computation of diagrams (i) and (ii) in Fig.~\eqref{OG3diags2PNh} yields: 
\begin{align}
L^{(h)}_{\rm (i)} & = \frac{4G^3 (m^0_1)^2 (m^0_2)^2 \alpha_1^0 \alpha_2^0}{r^3} \,,\\
L^{(h)}_{\rm (ii)} & = \frac{2G^3 (m^0_1)^2 (m^0_2)^2 (\alpha_1^0)^2 (\alpha_2^0)^2}{r^3} \,.
\end{align}

It is worthwhile mentioning here that, in all the computations carried out in this section, we have kept arbitrary dimensions and have worked in dimensional regularization, setting $d=3$ only at the final result for each diagram. This is important since the integrals \eqref{intKmaster} and \eqref{oneloopmasterintss} are naturally defined within the realm of dimensional regularization, and hence spurious divergences may appear in intermediate steps. Note also that, with the computation performed in this and the previous sections, we have explicitly checked that no ultraviolet nor infrared divergences arise up to the 2PN level.

\subsection{Results}

Finally, having computed all the relevant diagrams throughout this sections, we construct our 2PN Lagrangian for a compact binary system in ST theories by summing all these contributions to the 2PN GR Lagrangian in Eq.~(63) of Ref.~\cite{Gilmore:2008gq}, since the latter was obtained using precisely the same method and gauge choice, and which already include the kinetic term. We then remove all appearances of the sensitivities $\alpha_A, \beta_A, \beta'_A$ in favor of the ST parameters of Eqs.~\eqref{defG12}, \eqref{STparameters2}, and \eqref{STparameters3}, to write       
\begin{align}
L_{\rm 2PN} &= \frac{1}{16} m^0_1 \V_1^2  
            +\frac{G_{12}m^0_1 m^0_2}{16r} \bigg\{ 3 \V_1^2\V_2^2 + 2(\V_1\cdot\V_2)^2-2 \V_1^2 (\RR\cdot\V_2)^2 + 3(\RR\cdot\V_1)^2(\RR\cdot\V_2)^2 \nonumber\\
            &+2 (7+4 \bar{\gamma}_{12}) \V_1^4 -4 (5+2 \bar{\gamma}_{12}) \V_1^2 (\V_1\cdot\V_2)  +4 (3+2 \bar{\gamma}_{12}) (\RR\cdot\V_1)(\RR\cdot\V_2) (\V_1\cdot\V_2) \nonumber\\
            &-4 (3+2 \bar{\gamma}_{12}) (\RR\cdot\V_1)(\RR\cdot\V_2) \V_1^2 
            +2r [ \V_2^2(\RR\cdot\A_1) - (\RR\cdot\A_1) (\RR\cdot\V_2)^2 + 4 (3+2\bar{\gamma}_{12}) (\RR\cdot\V_2) (\V_1\cdot\A_1) \nonumber\\ 
            &- 2(7+4\bar{\gamma}_{12}) (\RR\cdot\V_2) (\V_2\cdot\A_1) ]  + r^2 \left[ (15+8\bar{\gamma}_{12}) (\A_1\cdot\A_2) - (\RR\cdot\A_1) (\RR\cdot\A_2) \right]\bigg\}  \nonumber\\
            &+ \frac{G_{12}^2 m^0_1 (m^0_2)^2}{4r^2} \bigg\{ 		
             \left( 7 + 10 \bar{\gamma}_{12} + \frac{7}{2} \bar{\gamma}_{12}^2 - 2 \delta_2 + 2\bar{\beta}_1 \right) \V_1^2 
            + 2 (4+ 6 \bar{\gamma}_{12} + 2 \bar{\gamma}_{12}^2 - \delta_2) \V_2^2 \nonumber\\
            &+ 2 \left( 1 + \bar{\gamma}_{12} + \frac{1}{4}\bar{\gamma}_{12}^2 + \delta_2 \right) (\RR\cdot\V_1)^2
             + 2 \left( -\bar{\gamma}_{12} - \frac{1}{4}\bar{\gamma}_{12}^2 + \delta_2 + 2 \bar{\beta}_1 \right) (\RR\cdot\V_2)^2 \nonumber\\
            &-2 \left( 7 + 11 \bar{\gamma}_{12} + \frac{15}{4} \bar{\gamma}_{12}^2 - 2\delta_2 \right) (\V_1\cdot\V_2) 
             -4 \delta_2 (\RR\cdot\V_1)(\RR\cdot\V_2)  -2r \left( \bar{\gamma}_{12} + \frac{1}{4} \bar{\gamma}_{12}^2 - 2 \bar{\beta}_1 \right) (\RR\cdot\A_2) \bigg\} \nonumber\\ 
            &+ \frac{G^3_{12} (m^0_1)^3 m^0_2}{2r^3} \left( 1 + \frac{2}{3}\bar{\gamma}_{12} + \frac{1}{6}\bar{\gamma}^2_{12} +2\bar{\beta}_2 + \frac{2}{3} \delta_1 + \frac{1}{3}\epsilon_2 \right) + \frac{G^3_{12} (m^0_1)^2 (m^0_2)^2}{2r^3} (3 + \bar{\gamma}_{12} + 4 \bar{\beta}_{1} + \zeta_{12}) \nonumber\\             
            &+ (1 \leftrightarrow 2)\,.
\end{align}
This is one of the most important results of the present paper, which, although different in form, is equivalent to the Lagrangian presented in Eq.~(B1c) of Ref.~\cite{Julie:2022qux} for the dynamics of binary system in ST theories, as discussed below. 
In particular, to the best of the author's knowledge, this is the first time this Lagrangian has been fully computed, and presented in detail, using EFT techniques. 
The relation between the two can be obtained by simply summing to the above result the following double-zero terms and total time derivatives\footnote{After publication of the present paper on the arXiv, one of the authors of Ref.~\cite{Bernard:2023eul} (S. Mougiakakos) brought to our attention that the same double-zero terms had also been obtained by them, and reported at Appendix B of their paper. Our results are in complete agreement with theirs.}
\begin{align}
\delta L_1 = &\frac{G_{12} m_1^0 m_2^0 r}{8} \bigg[ \left( \RR\cdot\A_1 + \frac{G_{12}m_2^0}{r^2} \right) \left( \RR\cdot\A_2 - \frac{G_{12}m_1^0}{r^2} \right) \nonumber \\
&-(15 + 8 \bar\gamma_{12}) \left( \A_1 + \frac{G_{12} m_2^0}{r^3}\R \right)\cdot\left( \A_2 - \frac{G_{12} m_1^0}{r^3}\R \right) \bigg]
\end{align}
and
\begin{align}
\delta L_2 = &\frac{1}{4}\frac{d}{dt} \bigg[  \left( 7 + 6 \bar\gamma_{12} + \frac{1}{2} \bar\gamma_{12}^2 - 4 \bar\beta_1 \right) \frac{G_{12}^2 m_1^0 (m_2^0)^2}{r} (\RR\cdot\V_2)  \nonumber \\
&+ (3 + 2 \bar\gamma_{12}) G_{12} m^0_1 m^0_2 (\RR\cdot\V_1) \V_2^2   \bigg] + (1 \leftrightarrow 2)\,.
\end{align}
Double zeros are harmless terms that can be added to the Lagrangian such that the equations of motion do not get modified (and hence providing a transformation between two equivalent Lagrangians), while staying in the same gauge. As it was shown long ago in Ref.~\cite{Barker:1980spx}, such terms are constructed out of the square of lower-order equations of motion. Notice that, simple replacement of the body's acceleration using the lower-order equations of motion can still be performed, but in this case such a transformation at the level of the Lagrangian modifies the gauge choice \cite{Schafer:1984mr}.

\section{The Gauss-Bonnet Contribution}\label{secGBcoupling0}

In this section, we study the effects of the Gauss-Bonnet coupling, present in Eq.~\eqref{phiaction0} through the coupling $\sim \alpha \sqrt{-g}f(\vphi)\mathcal{G}$, on the conservative dynamics of spinless compact binary systems. In this case, the new diagrams arising from such a coupling must be added to the results obtained above for ST theories in order to fully describe the conservative dynamics in EsGB gravity. Note that, the GB coupling introduces new scalar-graviton vertices, starting from a three-scalar-graviton-graviton vertex $\sim \Phi h h$, and defines an infinite ladder of scalar-graviton vertices comprising at least one scalar and two graviton modes.

\begin{figure*}
\centering
\begin{tikzpicture}
\begin{feynman}[scale=0.8, transform shape]
\vertex (a);
\vertex[small,dot,right=1.5cm of a] (b) {};
\vertex[small,dot,below=0.9cm of b] (g) {};
\vertex[below=0.45cm of b] (g0) {};
\vertex[right=1.5cm of g0] (g00);
\vertex[right=1.5cm of b] (c);
\vertex[below=1.8cm of a] (d);
\vertex[right=1.5cm of d] (e);
\vertex[small,dot,right=0.7cm of d] (e00) {};
\vertex[small,dot,right=2.3cm of d] (e000) {};
\vertex[right=1.55cm of e] (f);
\vertex[below=2.5cm of b] (y) {};
\node[shape=rectangle,fill=black] (n) at (1.55, -0.9) {\rule{0.005cm}{0.005cm}};
\vertex[right=1cm of c] (a2);
\vertex[small,dot,right=1.5cm of a2] (b2) {};
\vertex[small,dot,below=0.9cm of b2] (g2) {};
\vertex[below=0.45cm of b2] (g20) {};
\vertex[right=1.5cm of g20] (g200);
\vertex[right=1.5cm of b2] (c2);
\vertex[below=1.8cm of a2] (d2);
\vertex[right=1.5cm of d2] (e2);
\vertex[small,dot,right=0.7cm of d2] (e200) {};
\vertex[small,dot,right=2.3cm of d2] (e2000) {};
\vertex[right=1.55cm of e2] (f2);
\vertex[below=2.5cm of b2] (y2) {};
\node[shape=rectangle,fill=black] (n) at (5.6, -0.9) {\rule{0.005cm}{0.005cm}};
\diagram*{
(a) --[line width=0.8mm,plain] (c),
(a) --[line width=0.3mm,plain,white] (c),
(d) --[line width=0.8mm,plain] (f),
(d) --[line width=0.3mm,plain,white] (f),
(b) -- [\scalar] (g),
(g) -- [\gravphi] (e00),
(g) -- [\gravphi] (e000),
(a2) --[line width=0.8mm,plain] (c2),
(a2) --[line width=0.3mm,plain,white] (c2),
(d2) --[line width=0.8mm,plain] (f2),
(d2) --[line width=0.3mm,plain,white] (f2),
(b2) -- [\gravphi] (g2),
(e200) -- [\scalar] (g2),
(g2) -- [\gravphi] (e2000),
};
\vertex[dot,right=1.475cm of a] (bx) {};
\vertex[dot,right=0.675cm of d] (e00) {};
\vertex[dot,right=2.275cm of d] (e000) {};
\vertex[dot,below=0.9cm of bx] (g) {};
\vertex[above=0.4cm of bx] (v2) {$v^0$};
\vertex[below=0.4cm of e00] (v2) {$v^0$};
\vertex[below=0.4cm of e000] (v2) {$v^0$};
\vertex[right=0.45cm of g] (z) {};
\vertex[above=0.12cm of z] (v) {$v^0$};
\vertex[dot,right=1.475cm of a2] (b2) {};
\vertex[dot,right=0.675cm of d2] (e200) {};
\vertex[dot,right=2.275cm of d2] (e2000) {};
\vertex[dot,below=0.9cm of b2] (g2) {};
\vertex[above=0.4cm of b2] (v2) {$v^0$};
\vertex[below=0.4cm of e200] (v2) {$v^0$};
\vertex[below=0.4cm of e2000] (v2) {$v^0$};
\vertex[right=0.45cm of g2] (z2) {};
\vertex[above=0.12cm of z2] (v2) {$v^0$};
\end{feynman}
\begin{feynman}
\vertex[below=2.3cm of b] (bx) {(a)};
\vertex[below=2.3cm of b2] (bx) {(b)};
\end{feynman}
\end{tikzpicture}
\caption{Leading-order diagrams for the Gauss-Bonnet coupling. Here, we represent scalar-graviton vertices stemming from this new coupling by a square.}
\label{fig:diagGBcontribution}
\end{figure*}
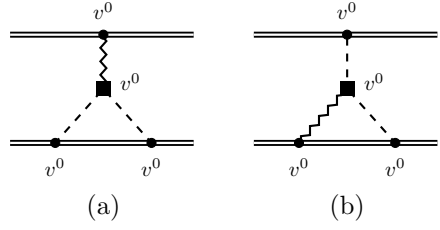

As the starting point, let us consider the leading-order contributions in the PN expansion, determined by the two diagrams in Fig.~\ref{fig:diagGBcontribution}, in which only $v^0$ vertices are considered. 
In this case, just the $\phi$ gravitational mode in the point-particle action is needed, as well as the $\Phi h h\rightarrow \Phi\phi\phi$ sector of the three-scalar-graviton-graviton vertex. Moreover, time derivatives can be neglected at this level, since, remember, they lead to higher-order PN corrections. Under these considerations, the three vertex in spacetime domain becomes very simple, and reads in arbitrary dimensions
\begin{equation}
S_{\Phi\phi\phi} \rightarrow \frac{4c_d \alpha f'(\vphi_0)}{\Lambda} (\delta_{ac}\delta_{bd}-\delta_{ab} \delta_{cd}) \int d^{d+1}x\, \Phi \partial_a\partial_b\phi \partial_c\partial_d\phi\,,
\end{equation}
where $f'(\vphi_0) \equiv (df/d\vphi)|_{\vphi\rightarrow\vphi_0}$.

Computation of diagrams (a) and (b) of Fig.~\ref{fig:diagGBcontribution} follows straightforwardly and results in
\begin{align}
L^{\rm GB}_{(a)} &= \frac{\alpha f'(\vphi_0)}{r^2} \frac{G^2 m_1^0 m_2^0}{r^2} (m_1^0 \alpha_2^0+m_2^0 \alpha_1^0) \,,\\
L^{\rm GB}_{(b)} &= \frac{2\alpha f'(\vphi_0)}{r^2} \frac{G^2 m_1^0 m_2^0}{r^2} (m_1^0 \alpha_1^0+m_2^0 \alpha_2^0) \,,
\end{align}
with total Lagrangian given by
\begin{equation}\label{GBcontr}
L^{\rm GB} = \frac{\alpha f'(\vphi_0)}{r^2} \frac{G^2 m_1^0 m_2^0}{r^2} \left[  m_1^0(\alpha_2^0 + 2 \alpha_1^0) + m_2^0(\alpha_1^0 + 2 \alpha_2^0) \right]\,.
\end{equation}
This result was originally derived in Ref.~\cite{BertiJulie2019}, and it is now computed here for the first time using EFT methods. 

Note that the inclusion of the Gauss-Bonnet term introduces a new length scale to the problem, that, in principle, has a distinct scaling in the PN approximation. In particular, the recent observations of gravitational waves, throughout the first two GW catalogs, have been used to put stringent constraints on the value of the coupling constant, placing the bounds of $\sqrt{\alpha} \lesssim 1.7$km \cite{Nair:2019iur,Perkins:2021mhb,Wang:2021jfc}. Because of this, it has been extensively considered in the literature the so-called small-$\alpha$ approximation $\alpha f'(\vphi) \lesssim (G M^0)^2$, where $M^0 = m_1^0 + m_2^0$ is the total asymptotic mass of the system; See also Ref.~\cite{Sotiriou_2014b} for similar theoretical constraints in the particular case of shift-symmetric scalar-Gauss-Bonnet gravity.

Now, rewriting the total Gauss-Bonnet contribution given in Eq.~\eqref{GBcontr} as
\begin{equation}
L^{\rm GB} = \frac{\alpha f'(\vphi_0)}{(G M^0)^2} \left(\frac{G M^0}{r}\right)^2 \frac{G^2 m_1^0 m_2^0}{r^2} \left[  m_1^0(\alpha_2^0 + 2 \alpha_1^0) + m_2^0(\alpha_1^0 + 2 \alpha_2^0) \right]\,,
\end{equation}
we see that, as discussed in Ref.~\cite{BertiJulie2019}, in the small-$\alpha$ limit, this term scales as a 3PN order correction. Therefore, although formally a 1PN correction, observational bounds push the leading-order correction from the Gauss-Bonnet coupling to start contributing just at the 3PN order.

Simple analysis of the $\mathcal{O}(G^3)$ diagram topologies of Fig.~\ref{fig:topologyG3} show us a similar behavior: for the diagrams in which just a single three vertex from the Gauss-Bonnet coupling is present, like in topologies (c)-(i), we have
\begin{equation}
L^{\rm 1-GB} \sim \frac{\alpha f'(\vphi_0)}{(G M^0)^2} \left(\frac{G M^0}{r}\right)^2 \frac{G^3 (m_1^0)^{3-p} (m_2^0)^p}{r^3} m_A F[\alpha_A,\beta_A,\dots]\,,
\end{equation}
for some integer $p$ and some function $F$ of the BH sensitivities. Hence, we easily see that, in the small-$\alpha$ limit, such contributions are equivalent to a 4PN order correction. In the case of diagrams which admit the insertion of two Gauss-Bonnet couplings, like diags.~(g)-(i), we rather have
\begin{equation}
L^{\rm 2-GB} \sim \left( \frac{\alpha f'(\vphi_0)}{(G M^0)^2}\right)^2 \left(\frac{G M^0}{r}\right)^4 \frac{G^3 (m_1^0)^{3-p} (m_2^0)^p}{r^3} m_A F[\alpha_A,\beta_A,\dots]\,,
\end{equation}
which, in turn, start contributing just at the 6PN order.

\section{Conclusion and Perspectives}\label{secconclusions0}

In this paper we have derived in detail the conservative dynamics of a gravitationally bound binary system in scalar-tensor and Einstein-scalar-Gauss-Bonnet gravity to the second post-Newtonian order for spinless objects. To this end, we have extended the effective field theory approach developed by Goldberger and Rothstein to include the new scalar degrees of freedom, which, as we have seen, couples minimally to gravity in the case of scalar-tensor theories and nonminimally in the case of the Einstein-scalar-Gauss-Bonnet gravity. In this derivation, we have shown how we systematically determine all the Feynman diagrams needed for each particular order, and then applied separately to the 0PN (Newtonian), 1PN, and 2PN cases in such theories. In particular, at the 2PN order, we encountered two-loop integrals that could always be reduced to the simple one-loop ones found at the 1PN level. In all the cases, the diagrams could be computed by using just two master integrals, thus further demonstrating the efficiency of the method.

Computations were carried out using the Kaluza-Klein parametrization, which, as extensively demonstrated in the pure GR case, it yields a substantial reduction in the number of diagrams as compared to when the standard Lorentz covariant metric parametrization is used. Here, we have explicitly demonstrated that, up to the 2PN level, this also happens in scalar-tensor theories. Note that, while at the 0PN and 1PN orders the number of additional diagrams due to the scalar interactions is equal to the number of diagrams originally present in GR (one for the 0PN and four for the 1PN), at the 2PN order we have 34 new diagrams that must be added to the 21 ones of GR. In particular, we expect this grow in diagrams, easily understood as having origin in the new possibilities of interactions, to become even more severe as we consider higher PN orders, occasion on which the Kaluza-Klein parametrization becomes decisive.

Finally, the results derived in this paper provide an essential step toward the correct description of the conservative dynamics of compact binaries systems at the 3PN level in ST and EsGB theories. Recall that, as discussed by Julié and Berti, the 3PN equations of motion obtained by Bernard via traditional methods present inconsistencies in some coefficients, making EFT methods timely as it provides an excellent complementary, and independent, way of comparing results. Note also that, going for the 3PN dynamics is crucial for the modeling, and possible further detection, of effects stemming from the Gauss-Bonnet coupling, since, as it was shown in this paper using EFT methods, this is the order in which such effects first appear. Nevertheless, it is worthwhile pointing out that, although these are still not present at the 2PN order, the influence of the Gauss-Bonnet coupling will be imprinted in the BH sensitivities. In any case, observation of deviations from GR in binary BH coalescences that are compatible with ST theories will certainly indicate the existence of hairy black holes in our Universe, although not necessarily from EsGB gravity, since many other beyond GR theories that admit hairy solutions have also ST theories as a particular case.

\section*{Acknowledgments}

GLA would like to thank the Yukawa Institute for Theoretical Physics at the University of Kyoto (while in participation in the Gravity and Cosmology 2024 long-term workshop) for the hospitality during the final stages of this work. The work of GLA is supported by the National Natural Science Foundation of China under Grant No. 12247103.

\bibliography{bibliography}

\end{document}